\documentclass[a4paper,fleqn,usenatbib]{mnras}

\usepackage[T1]{fontenc}
\usepackage{ae,aecompl}

\usepackage{graphicx}	% Including figure files
\usepackage{amsmath}	% Advanced maths commands
\usepackage{amssymb}	% Extra maths symbols
\usepackage{soul}

\title[Detection of O II in Type Ia SN 2010kg]{Possible detection of singly-ionized oxygen in the Type Ia SN 2010kg}

\author[B. Barna et al.]{
B. Barna$^{1}$,
J. Vinko$^{1,2}$,
J. M. Silverman$^{2,3}$,
G. H. Marion$^{2}$,
J. C. Wheeler$^{2}$
\\
$^{1}$Department of Optics and Quantum Electronics, University of Szeged, Dom ter 9 Szeged, Hungary\\
$^{2}$Department of Astronomy, University of Texas at Austin, 1 University Station C1400, Austin, TX 78712-0259, USA\\
$^{3}$NSF Astronomy and Astrophysics Postdoctoral Fellow
}

\date{Accepted XXX. Received YYY; in original form ZZZ}

\pubyear{2015}

\begin{document}
\label{firstpage}
\pagerange{\pageref{firstpage}--\pageref{lastpage}}
\maketitle

% Abstract of the paper
\begin{abstract}

We present direct spectroscopic modeling of 11 high-S/N observed spectra of the Type Ia SN 2010kg, taken between -10 and +5 days with respect to B-maximum. The synthetic spectra, calculated with the SYN++ code, span the range between 4100 and 8500 \r{A}. Our results are in good agreement with previous findings for other Type Ia SNe. Most of the spectral features are formed at or close to the photosphere, but some ions, like \ion{Fe}{ii} and \ion{Mg}{ii}, also form features at $\sim$2000 - 5000 km s$^{-1}$ above the photosphere. The well-known high-velocity features of the \ion{Ca}{ii} IR-triplet as well as \ion{Si}{ii} $\lambda$6355 are also detected.

The single absorption feature at $\sim$4400 \r{A}, which usually has been identified as due to \ion{Si}{iii}, is poorly fit with \ion{Si}{iii} in SN 2010kg. We find that the fit can be improved by assuming that this feature is due to either \ion{C}{iii} or \ion{O}{ii}, located in the outermost part of the ejecta, $\sim$4000 - 5000 km s$^{-1}$ above the photosphere.
Since the presence of \ion{C}{iii} is unlikely, because of the lack of the necessary excitation/ionization conditions in the outer ejecta, we identify this feature as due to \ion{O}{ii}. The simultaneous presence of \ion{O}{i} and \ion{O}{ii} is in good agreement with the optical depth calculations and the temperature distribution in the ejecta of SN 2010kg. This could be the first identification of singly ionized oxygen in a Type Ia SN atmosphere.

\end{abstract}

% Select between one and six entries from the list of approved keywords.
% Don't make up new ones.

%%%%%%%%%%%%%%%%%%%%%%%%%%%%%%%%%%%%%%%%%%%%%%%%%%

%%%%%%%%%%%%%%%%% BODY OF PAPER %%%%%%%%%%%%%%%%%%

\section{Introduction}

The model of Type Ia supernovae (SNe Ia) is widely accepted to be a thermonuclear explosion of a C-O white dwarf \citep{hoyle1960}, although there are several open questions about the progenitor system and the details of explosion. The cosmological application of Ia SNe is based on the standardization of their luminosity \citep{matheson2012} because their peak luminosities are well correlated with the decline rates ($\Delta m_{15}$(\textit{B})) of their light curves \citep{phillips1993}.
In order to increase the effectiveness of Ia SNe as distance indicators, we have to find the connection between the observed diversity of Ia SNe \citep{branch2006, blondin2012} and the different explosion models and/or combustion methods. The two frequently considered explosion channels are the single-degenerate \cite[the WD accretes material from its non-degenerate companion;][]{whelan1973} and the double-degenerate \cite[the progenitor merges with another WD;][]{iben1984, webbink1984} model. The burning propagation can occur as deflagration \cite[propagation with subsonic speed;][]{nomoto1984} or delayed detonation \cite[subsonic propagation turns into supersonic;][]{khokhlov1991}. The study of the early spectra and the investigation of the ion signatures may lead us to the explanation of the Ia SNe diversity.

\begin{table*}
	\centering
	\caption{Journal of the spectroscopic observations. The columns contain the followings: date of observation, modified Julian date, phase with respect to the moment of maximum light in B-band, exposure time, spectral range, spectral resolution in \r{A} and signal-to-noise ratio, respectively.}
	\label{tab:table1}
	\begin{tabular}{l|ccccccc} % four columns, alignment for each
		\hline
		 Date & MJD & Phase (days) & Exp. time (s) & Spectral range (\r{A}) & $\Delta \lambda$ (\r{A})& S/N\\
		\hline
		2010-12-02 & 55532.201  & -10  &  1200 &	4100-10,500	&  19 &    77\\
		2010-12-04 & 55534.196  &  -8  &  1100 &	4100-10,500	&  19 &   110\\
		2010-12-06 & 55536.190  &  -6  &   900 &	4100-10,500	&  19 &   108\\
		2010-12-07 & 55537.190  &  -5  &   900 &	4100-10,500	&  19 &   117\\
		2010-12-08 & 55538.183  &  -4  &   700 &	4100-10,500	&  19 &    98\\
		2010-12-09 & 55539.193  &  -3  &   700 &	4100-10,500	&  19 &   104\\
		2010-12-11 & 55541.174  &  -1  &  2235 &	4100-10,500	&  19 &   106\\
		2010-12-12 & 55542.174  &   0  &   600 &	4100-10,500	&  19 &   113\\
		2010-12-13 & 55543.170  &  +1  &   600 &	4100-10,500	&  19 &    79\\
		2010-12-15 & 55545.164  &  +3  &   600 &	4100-10,500	&  19 &    83\\
		2010-12-17 & 55547.167  &  +5  &   600 &	4100-10,500	&  19 &    90\\
		\hline
	\end{tabular}
\end{table*}

In this paper, we present the direct analysis of eleven spectra of the Type Ia SN 2010kg. After the overview of our dataset in Section~\ref{sec:dataset}, our spectrum fitting method is presented, focusing on the ion identification and the limitation of the approximations used by SYN++ \citep{thomas2011} in Section~\ref{sec:methods}. The results of the fitting process are presented and discussed in Section~\ref{sec:results}, while in Section~\ref{sec:discussion} we focus on the temperature profile of SN 2010kg and the possible presence of unburnt material in the outer ejecta. Finally, we summarize our conclusions in Section~\ref{sec:conclusions}.

\section{Dataset}
\label{sec:dataset}

SN 2010kg was discovered on 30th November 2010 \citep{nayak2010}, by the Lick Observatory Supernova Search team with the Katzman Automatic Imaging Telescope. The supernova exploded in NGC 1633, with redshift of z = 0.0166 \citep{springob2005}. From the spectroscopic measurements, it was identified as a broad-line Type Ia supernova at more than 1 week before maximum brightness \citep{marion2010, silverman2010}.

Broad-line (BL) SNe are a subclass of Type Ia, defined by \cite{branch2006}, based on the strength of the
\ion{Si}{ii} $\lambda$6355 \r{A} feature, which is much broader than in core-normal SNe Ia. As former studies showed \citep{blondin2013, blondin2015}, broad \ion{Si}{ii} $\lambda$6355 lines in Type Ia SNe could be the result of delayed detonations. The spectra usually display high mean expansion velocities for all line-forming regions, while the decline rate of the velocities show large variation \citep{blondin2012, silverman2012a}. 

Our data sample consists of eleven spectra of SN 2010kg between -10 and +5 days with
respect to the moment of maximum luminosity in B-band (Dec 12.5, 2010). All of them were obtained with the Marcario Low Resolution Spectrograph attached to the Hobby-Eberly Telescope (HET) at McDonald Observatory, Texas \citep{hill1998}. All spectra extend from 4,100 to 10,500 \r{A} with resolution of $\sim$19 \r{A} and have signal-to-noise ratio between 70 and 120 (see Table~\ref{tab:table1}). Before the analysis, all spectra were corrected for redshift and the Milky Way interstellar reddening, assuming $R_V$ = 3.1 and the extinction curve of \cite{fitzpatrick2007}, based on dust emission maps from COBE/DIRBE and IRAS/ISSA \citep{schlafly2011}. We applied $E(B-V)$ = 0.137 mag for dereddening. Due to the cut-off at 4100 \r{A} in the HET spectra, we are able to analyse some of the strongest features, such as the \ion{Ca}{ii} near-infrared triplet and \ion{Si}{ii} $\lambda$6355, but not the \ion{Ca}{ii} H$\&$K lines. Since the flux calibration becomes unreliable near the edges of the observed spectra, we restrict our analysis to the spectral range between 4300 and 8500 \r{A}.

\section{Methods}
\label{sec:methods}

\subsection{SYN++}
\label{sec:syn++}

We use the SYN++ code \citep{thomas2011}, which is the modern version of the original SYNOW spectrum synthesis code \citep{fisher1997} rewritten in C++. The computation follows the Sobolev-approximation \citep{sobolev1960}: the blackbody radiation emitted by the sharp, spherical photosphere interacts with the expanding atmosphere, which has a large velocity gradient and low density. The main simplification is that a photon interacts with an atom only in a narrow region within the ejecta, where it suffers resonance scattering. The result is a P Cygni profile, where the pseudo-absorption is blueshifted with respect to the line central wavelength.

The global input parameters for the whole spectrum model are the velocity (hereafter $v_{phot}$) and the temperature ($T_{phot}$) of the photosphere. SYN++ approximates the photosphere as a sharp, spherically symmetric surface, which emits pure blackbody flux and its blackbody temperature is $T_{phot}$. Although the shape of the SN spectra can be more-or-less described as a diluted blackbody  \cite[e.g.][]{jeffery1990}, LTE conditions probably never occur in Type Ia ejecta, thus, the derived $T_{phot}$ values may not represent the actual local temperatures (see also Sec. \ref{sec:phottemperature}).

Since SYN++ computes relative fluxes, we have to multiply the absolute fluxes of the observed spectra with a flux scaling parameter ($a0$ in the input file of the code), which also needs to be set for each spectrum. SYN++ is also able to scale the spectrum with a linear and quadratic function of the wavelength, but we set these warping parameters ($a1$ and $a2$ in the input file) to zero in order to reproduce the shape of the blackbody continuum without additional distortion. There are additional parameters for every ion: the optical depth ($\tau$), the minimum velocity of the line-forming region (hereafter minimum velocity or $v_{min}$), the e-folding velocity of the opacity profile ($aux$ in the input file) and the Boltzmann excitation temperature ($T_{exc}$). We fix the maximum velocity for all line-forming regions at $v_{max}$ = 40,000 km s$^{-1}$. All these ion-parameters refer to a reference line of the given ion. The optical depths of every other lines are calculated based on the $T_{exc}$ parameter and the atomic data included by SYN++. The applied atomic data were downloaded from the homepage of \textit{ES: Elementary Supernova Spectrum Synthesis User Guide}\footnote{https://c3.lbl.gov/es/}, which is the enhanced version of database of \cite{kurucz1993}.

The fitting code, SYNAPPS, varies the above mentioned parameters automatically and simultaneously, and fits the model to the data via $\chi^{2}$-minimisation. A caveat in such a complex fitting is that the large number of parameters may make the final best-fit model ambiguous, because the fitting algorithm could stick in an incorrect parameter combination due to the strong correlations between the parameters. With this in our view, we used mainly the SYN++ code instead of SYNAPPS and varied the input parameters interactively to get better fits.

\begin{figure*}
	% To include a figure from a file named example.*
	% Allowable file formats are eps or ps if compiling using latex
	% or pdf, png, jpg if compiling using pdflatex
	\includegraphics[width=160mm]{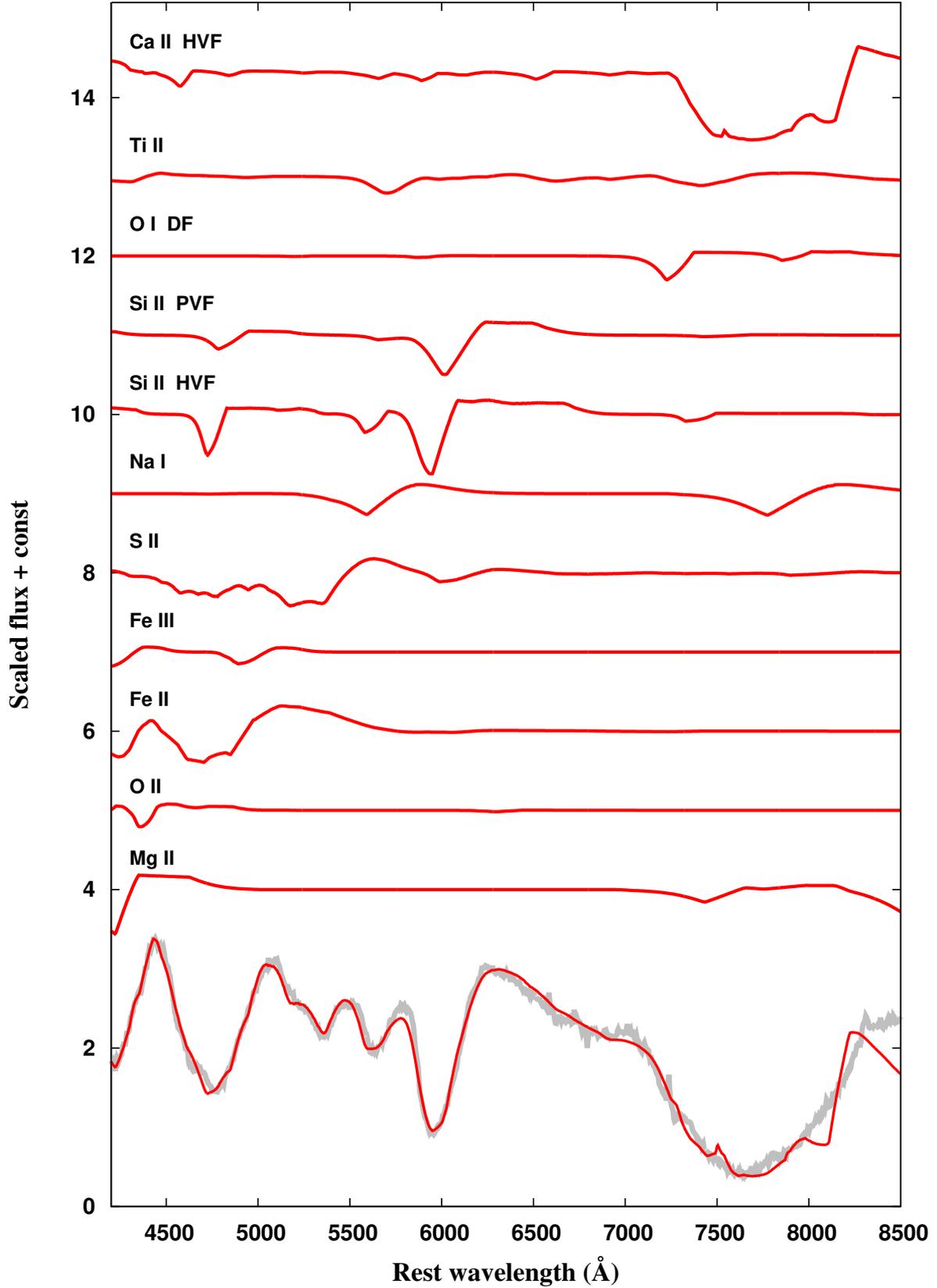}
    \caption{The single-ion synthetic spectra ten days before maximum light in SN 2010kg; the observed spectrum (grey line) and the whole synthetic spectrum (red line)}
    \label{fig:fig1}
\end{figure*}

\begin{figure*}
	% To include a figure from a file named example.*
	% Allowable file formats are eps or ps if compiling using latex
	% or pdf, png, jpg if compiling using pdflatex
	\includegraphics[width=160mm]{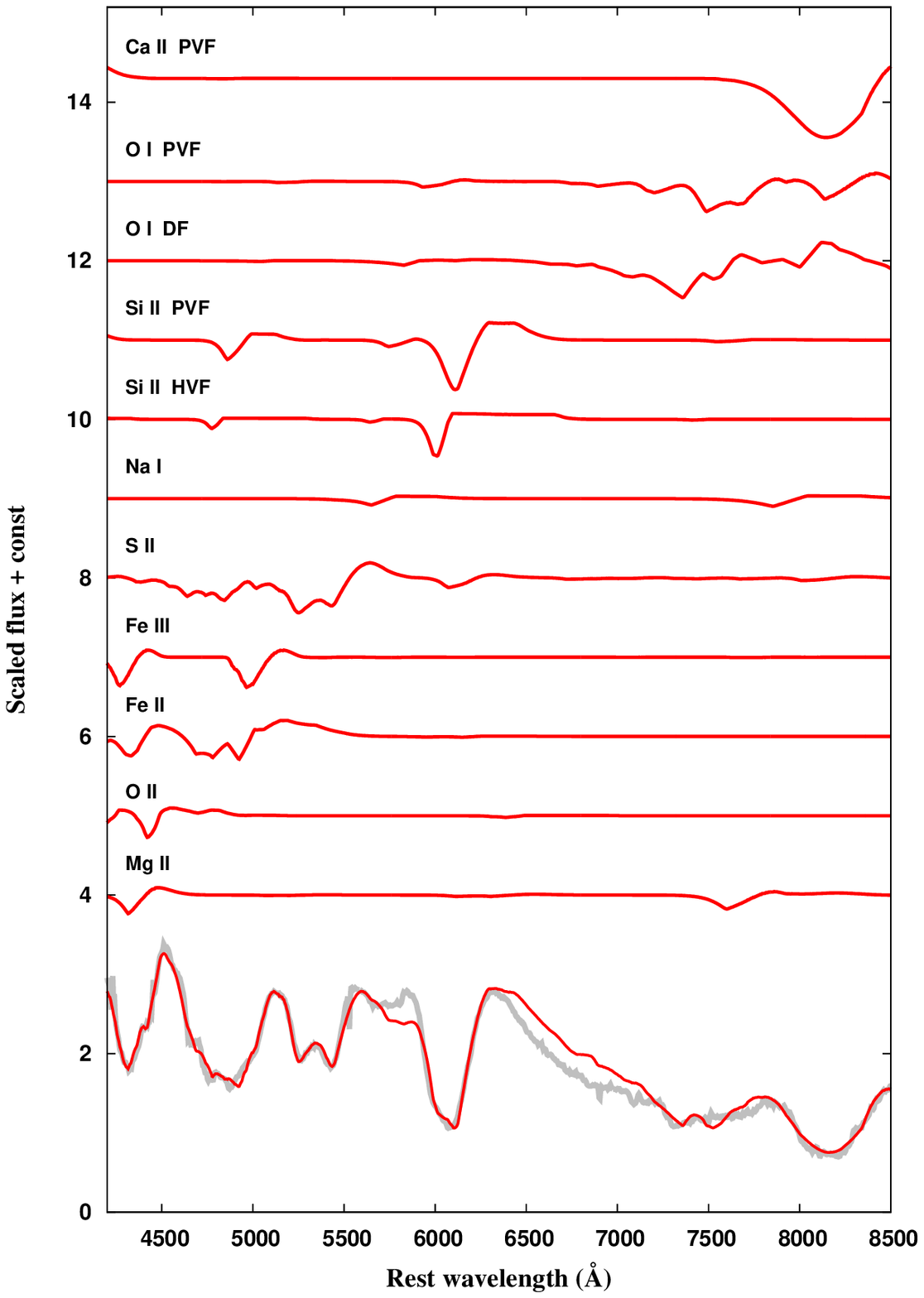}
    \caption{The single-ion synthetic spectra at the maximum light in SN 2010kg; the observed spectrum (grey line) and the whole synthetic spectrum (red line)}
    \label{fig:fig2}
\end{figure*}

\subsection{Ion identification}
\label{sec:ions} % used for referring to this section from elsewhere

During the photospheric phase, the P Cygni line profiles are dominant in the Type Ia SN spectrum. We focus on the spectra taken earlier than +14 days post-maximum, because the physical assumption for the line forming mechanism (pure resonant scattering) in the SYN++ code may be more-or-less valid roughly before this epoch. At the moment of maximum light, the \ion{Ca}{ii} H$\&$k lines at $\lambda3934$ and $\lambda3969$, the "W" feature caused by \ion{S}{ii} at $\sim$5500 \r{A}, the \ion{Si}{ii} $\lambda$6355 and the \ion{Ca}{ii} NIR triplet are always observable.

Most of the lines in the supernova spectrum overlap each other due to the fast expansion of the ejecta, making the line identification always ambiguous. Owing to the several independent studies \cite[e.g.][]{branch2006, tanaka2008, parrent2012, hsiao2013, marion2015, jack2015}, we can assume the existence of a few types of ions in the ejecta. These are typically \ion{Si}{ii}, \ion{S}{ii}, \ion{Ca}{ii} (as mentioned above) supplemented by \ion{Fe}{ii}, \ion{Fe}{iii}, \ion{Mg}{ii} and \ion{O}{i}. Some studies pointed out the possible contribution of \ion{Na}{i} \citep{branch2005}, \ion{C}{i} \citep{hsiao2015}, \ion{C}{ii} \citep{silverman2012b}, \ion{Ti}{ii} \citep{filippenko1992}, \ion{Ni}{ii} and \ion{Co}{ii} \citep{branch2005} at least in a fraction of Ia spectra taken before maximum light.

We performed the ion identification based on the literature mentioned above, and the detailed optical depth calculations presented by \cite{hatano1999a}. The reliably identified ions are \ion{Ca}{ii}, \ion{Si}{ii}, \ion{S}{ii}, \ion{Fe}{ii}, \ion{Fe}{iii}, \ion{O}{i} and \ion{Mg}{ii}.
Furthermore, we use \ion{Na}{i} and \ion{Ti}{ii} for some synthetic spectra to get better fitting, but their existence - without a strong absorption peak - remains ambiguous. Contribution from \ion{Ni}{ii} and \ion{Co}{ii} are also tested in order to reproduce the strong flux decrease below 4000 \r{A} observed in all Type Ia SNe \citep{branch1986, foley2012a}. The contributions from all the individual ions at two different epochs are presented in Fig. \ref{fig:fig1} and \ref{fig:fig2}.

An open question concerning the line identification in Type Ia SNe is the origin of the deep, narrow absorption feature at $\sim$4400 \r{A}. Looking through the literature, many authors fit this region with \ion{SI}{iii} \cite[e.g.][]{marion2014, yamanaka2009} and with a mixed contribution of \ion{Co}{ii} and \ion{Fe}{ii}. We find that, at least in the case of SN 2010kg, either \ion{C}{iii} or \ion{O}{ii} may give a better fit than \ion{Si}{iii}; in Section~\ref{sec:oii} we discuss the possible absence / presence of \ion{C}{iii} / \ion{O}{ii}, respectively.

\begin{figure*}
	% To include a figure from a file named example.*
	% Allowable file formats are eps or ps if compiling using latex
	% or pdf, png, jpg if compiling using pdflatex
	\includegraphics[width=130mm]{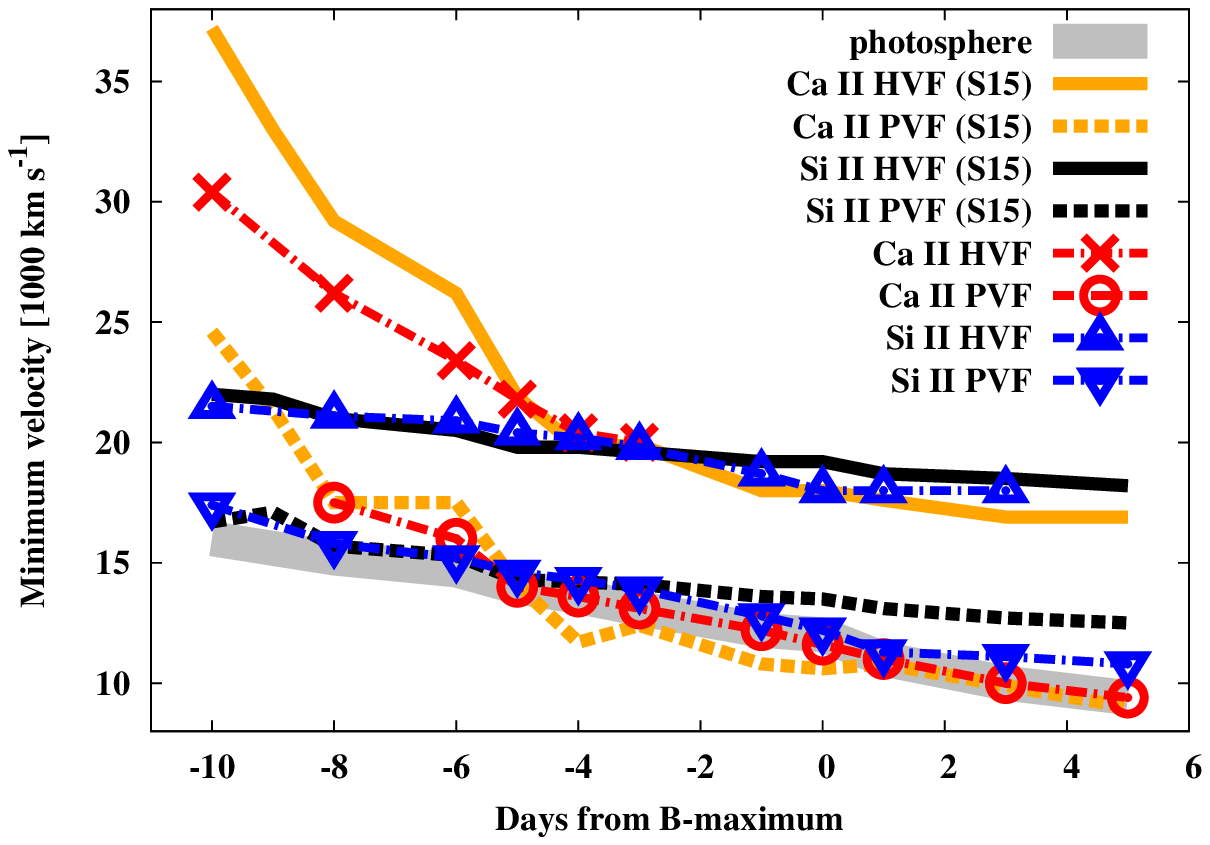}
    \caption{The velocities of the high-velocity features and their photospheric components in SN 2010kg, compared to the results for SN 2010kg of \protect\cite{silverman2015}. The photospheric velocity is presented with grey solid line.}
    \label{fig:fig3}
\end{figure*}

\subsection{High Velocity Features}
\label{sec:hvf}

High Velocity Features (HVF) have been the focus of interest since their first detection \citep{hatano1999b}. Those ions, with $v_{min}$ being $\sim$7000 - 15,000 km s$^{-1}$ higher than the photospheric velocity ($v_{phot}$), produce more highly blueshifted absorption features \cite[e.g. Silverman et al. 2015, hereafter][]{silverman2015,mazzali2005}. Together with the second component formed close to the photosphere (Photospheric Velocity Feature, PVF), they produce a typical double-bottom line profile. In homologously expanding ejecta, the different velocity values indicate that the line-forming regions of HVF and PVF are physically distinct, which is supported by the spectropolarimetric observations \citep{maund2013}. The origin of HVFs are still uncertain; density or abundance enhancement, caused by swept-up gas \citep{gerardy2004} or clumpy circumstellar material (CSM) \citep{kasen2003}, could explain their origin. 

After fifteen years of research it seems certain that \ion{Ca}{ii} HVFs appear in most of the SNe Ia ejecta at early epochs and they become weaker as the SN evolves. The only exception is the group of the peculiar 91bg-like SNe Ia where almost no HVFs are seen \citep{silverman2015}. \ion{Si}{ii} HVFs are more rare. Approaching maximum light, \ion{Si}{ii} HVF usually disappear, and only \ion{Ca}{ii} HVF survives the  maximum \citep{childress2014}. Some studies have also shown that other ions, like \ion{Fe}{ii} \citep{marion2014} or \ion{O}{i} \citep{parrent2012}, can produce HVFs a few days after the explosion. Unfortunately, the broad, overlapping lines can easily blend with these features, thus the identification of these HVFs in the case of other ions is still ambiguous in SN 2010kg. In our study, we investigate all the ions at higher velocities to create constraints on the real HVFs. 

Nevertheless, the situation (and thus the whole spectral modelling) is more complex because of the possible presence of detached features \cite[DF,][]{jeffery1990}. Their line-forming regions are above the photosphere, so the produced DFs show higher velocities than $v_{phot}$, but these velocities stay below the $v_{min}$ of HVFs. Thus, unlike the HVFs, the origin of DFs is unlikely to be related to the presence of CSM. The minimum velocity of a DF is at least $\sim$1500 - 2000 km s$^{-1}$ higher than $v_{phot}$ at any epochs and roughly lower than $\sim$19 - 20,000 km s$^{-1}$ one week before maximum light. Note that the upper velocity limit of the DF region depends on the epoch and cannot be fixed to an exact velocity value. Thus, we identify the features with $v_{phot}$ < $v_{min}$ < $v_{phot}$ + 2000 km s$^{-1}$  as photospheric velocity features (PVFs), while those having $v_{min}$ > $v_{phot}$ + 6000 km s$^{-1}$ as HVFs. In between these two regimes, the features are considered as DFs.

Since the origin of HVFs is uncertain, and they might be formed outside the SN ejecta (e.g. due to CSM-interaction), we refer the line forming region of the DFs with highest $v_{min}$ (with $\sim$5 - 6000 km s$^{-1}$ above the photosphere) as the outer region of the SN ejecta hereafter. Based on the recent SN Ia models \cite[e.g.][]{dessart2014c,seitenzahl2013}, oxygen and carbon DFs may be expected, because of their higher mass fraction in the outer layers ($v_{min}$ > 16,000 km s$^{-1}$), , but these are also somewhat model dependent because of the strength of mixing involved in the models \cite[see e.g. Fig.2 of][]{dessart2014b}.

\section{Results}
\label{sec:results}

Adopting the velocity values from the fits, we are able roughly to map the spatial distribution of the ions in the ejecta. This can help us getting abundance information and creating constraints on the explosion mechanism. In the following, we discuss each individual ion identified in the spectra of SN 2010kg.

\begin{figure*}
	% To include a figure from a file named example.*
	% Allowable file formats are eps or ps if compiling using latex
	% or pdf, png, jpg if compiling using pdflatex
	\includegraphics[width=130mm]{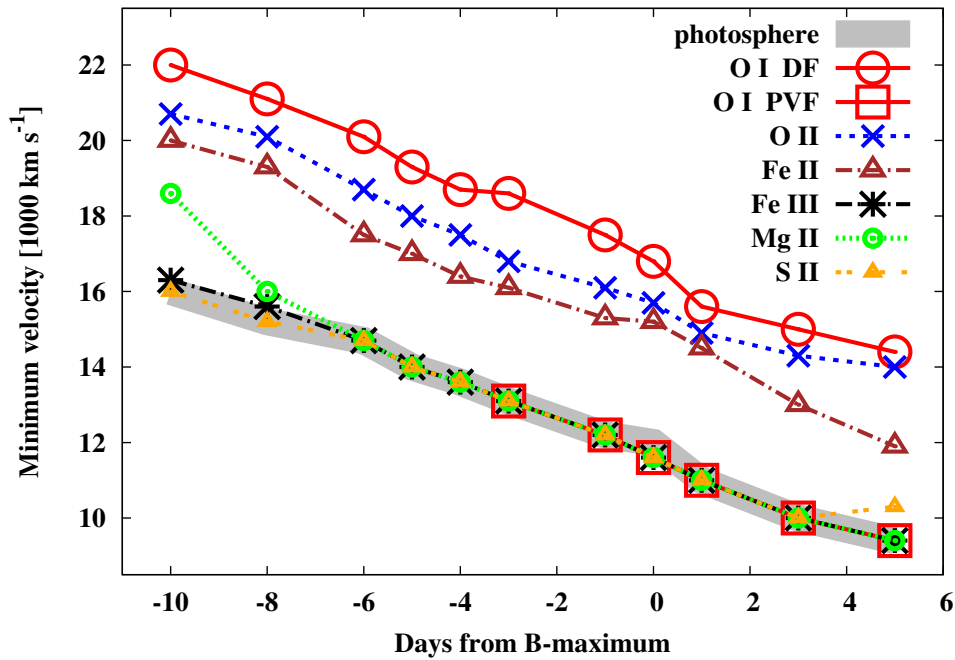}
    \caption{The velocities of ions with photospheric or detached line forming regions. The photospheric velocity is presented with the thick solid grey line, while the \ion{Si}{ii} HVF velocity (as the assumed upper limit of the SN ejecta) is plotted with the thin grey line.}
    \label{fig:fig4}
\end{figure*}

\subsection{Ca II lines}
\label{sec:caII}

The singly-ionized calcium produces the strongest HVFs and the highest minimum velocities in Type Ia SNe spectra. Several studies \cite[e.g.][]{childress2014, silverman2015} showed that the \ion{Ca}{ii} HVFs can survive even after maximum light. These characteristics make the \ion{Ca}{ii} H$\&$K lines and the NIR triplet ideal for studying HVFs. The HET spectra used in this study do not cover the region below 4100 \r{A}, thus, \ion{Ca}{ii} H$\&$K lines cannot be examined. The NIR triplet is surpassingly strong compared to spectra of other Type Ia SNe. The triplet is built up by \ion{Ca}{ii} $\lambda8498$, $\lambda8542$ and $\lambda8662$. Although, because of the high optical depth values needed at the first epoch, even \ion{Ca}{ii} $\lambda8927$ appears at the red wing of the feature (at $\sim$8200 \r{A} in Fig. \ref{fig:fig1}), whose presence disturbs the fitting. No other significant line of \ion{Ca}{ii} appears in our synthetic spectra between 4100 - 8500 \r{A}.

The best-fit is made using two distinct line-forming regions during most of the pre-maximum epochs, but with a few exceptions. The PVF component does not seem necessary to fit the \ion{Ca}{ii} NIR feature at -10 days (see Fig. \ref{fig:fig1}). Since the \ion{Ca}{ii} triplet is a very wide feature, the \ion{Ca}{ii} PVF component strongly overlaps with its HVF after -4 days. However, SYN++ is not able to resolve such a complex spectral feature. When the optical depth profiles of the strong PVF and HVF components of \ion{Ca}{ii} starts to overlap in velocity space, SYN++ assumes that only the stronger component at a particular wavelength (instead of the sum of the two) takes part in the line-forming process. This modelling assumption is completely different from the method of \cite{childress2014} and \cite{silverman2015}, where the authors co-added the best-fit pseudo-absorptions of the PVF and HVF components (see below in this section).

With the lack of observed \ion{Ca}{ii} H$\&$K lines, we cannot create additional constraints on the \ion{Ca}{ii} HVF line forming region, and its properties become highly uncertain when the overlapping with the PVF component takes place. After -3 days, the PVF component shows a decent fit alone, and there is no direct sign of a resolved HVF component. Thus, we eliminated the \ion{Ca}{ii} HVF from our SYN++ models after this epoch. Note that the ignorance of \ion{Ca}{ii} HVF does not necessarily imply the physical disappearance of the HVF line forming layer (see Sec \ref{sec:hvf}). Similar result was presented by \cite{vallely2015} for 2014J, where \ion{Ca}{ii} HVF was found absent in the near- and post-maximum SYNOW models.

At -8 and -6 days the $v_{min}$ of the PVF component is slightly higher than $v_{phot}$, but later they run together (see Fig. \ref{fig:fig3}). By extrapolating the two earliest \ion{Ca}{ii} PVF velocities to -10 days, it is conceivable that the \ion{Ca}{ii} PVF starts as a DF shortly after the explosion, but because of its higher decline rate, later it decreases down to the photosphere, like other detached ions. The minimum velocity of the \ion{Ca}{ii} HVF starts at 30,400 km s$^{-1}$, but declines fast until 4 days before maximum light (see Fig. \ref{fig:fig3}). At the last two epochs (-4 and -3 days), when \ion{Ca}{ii} HVF is still detectable, the $v_{min}$ does not decline further, which may support the idea that \ion{Ca}{ii} HVF has a velocity floor in SN 2010kg  \cite[][]{silverman2015}.

In Fig. \ref{fig:fig3}, the minimum velocities for \ion{Ca}{ii} and \ion{Si}{ii} are compared to the results for SN 2010kg of \cite{silverman2015}. Those authors fit separate Gaussian profiles to the PVF and HVF components. They defined the PVF and HVF velocities as the Doppler shifts of the absorption minima of each Gaussian component, which are found to be consistent with the velocities from fitting P Cygni line profiles given by SYN++. As expected, the \ion{Ca}{ii} velocities from these two studies are in good agreement. The difference between the HVF-velocities at the epoch of -10 days is probably due to the fact that we applied only the HVF component of \ion{Ca}{ii} for fitting the entire feature at this phase, while \cite{silverman2015} used both the HVF and PVF components for fitting the same spectrum.

\subsection{Si II lines}
\label{sec:siII}

As for \ion{Ca}{ii}, the singly-ionized silicon shows both PVF and HVF. Although \ion{Si}{ii} forms another line at $\sim$4900 \r{A}, this region is strongly overlapped by \ion{Fe}{ii} features. As a result, the only \ion{Si}{ii} feature where the HVF velocity is measurable is the $\lambda$6355. The absorption components of the separate HVF + PVF can be seen by eye, just like the transition between the two components at 4 days before maximum light, after which the PVF becomes the dominant feature. Unlike the \ion{Ca}{ii} NIR triplet, there are no multiple lines of \ion{Si}{ii} in the region of $\lambda$6355, thus, the HVF and PVF components of this feature stay resolved for longer time (Fig. \ref{fig:fig2}).

The velocity of the photospheric component decreases similarly to $v_{phot}$, but it is always higher by $\sim$500 - 1500 km s$^{-1}$ than $v_{phot}$ (see Fig. \ref{fig:fig3}). These results show good agreement with the velocity values obtained by \cite{silverman2015}. The $v_{min}$ of \ion{Si}{ii} HVF does not change strongly, unlike the \ion{Ca}{ii} HVF, the decline rate is only 200 km s$^{-1}$ day$^{-1}$ on the average. After the B-band maximum, the \ion{Si}{ii} HVF does not disappear, contrary to most other SNe studied in recent literature. Three days after maximum light its contribution to the absorption is still significant (see Fig. \ref{fig:fig12}). At this time, its minimum velocity is $\sim$18,000 km s$^{-1}$, which is similar to the velocity floor of \ion{Ca}{ii} HVF in \cite{silverman2015}. Based on the high velocity floor of the HVFs in SN 2010kg, it seems possible that the origin of the HVFs might be related to some kind of interaction between the ejecta and an outer density enhancement or CSM.

\subsection{Oxygen lines}
\label{sec:oxy}

HVF of neutral oxygen is identified as the strongest \ion{O}{i} feature at $\lambda$7773 \citep{parrent2012, nugent2011}, thus it seems reasonable to set both a PVF and a detached component in our models. At the early epochs, the contribution of \ion{O}{i} PVF is negligible, thus we use only the DF component. This component shows the highest velocity among all the DFs in SN 2010kg (see Fig. \ref{fig:fig4}), which is even higher than the HVF of \ion{Si}{ii} at -10 days. Later, the line forming region of \ion{O}{i} sinks below the HVF line forming regions, but stays above any other ions, being at $\sim$5000 km s$^{-1}$ above the photosphere. Because of the continuous decreasing of its $v_{min}$, we rather identify \ion{O}{i} DF than HVF. The behaviour of \ion{O}{i} suggests that the transition between HVFs and DFs may be continuous, and their line forming regions may not be separated sharply.

At three days before maximum light, the \ion{O}{i} DF cannot fit the red side of the $\lambda$7773 feature alone, because the flux from the pseudo-emission of \ion{O}{i} DF is too high to fit this spectral region. The application of \ion{O}{i} PVF improves the fitting till the end of the observed epochs. The velocity of the PVF is bound to $v_{phot}$ in our SYN++ models (see Fig. \ref{fig:fig4}). However, the two components (DF and PVF) cannot be seen separate, unlike in the case of \ion{Si}{ii} $\lambda$6355. Thus, the detection of \ion{O}{i} PVF is ambiguous in SN 2010kg. Further testing of the simultaneous appearance of both the DF and the PVF components for \ion{O}{i} may be possible in the NIR domain, where more oxygen features can be identified \citep{marion2009}.

Note that we also apply ionized oxygen (\ion{O}{ii}) to achieve better fitting for the absorption feature at $\sim$4400 \r{A}. We discuss the possible presence of \ion{O}{ii} in the Section~\ref{sec:oii}.

\subsection{Other ions}
\label{sec:other}

Singly ionized magnesium (\ion{Mg}{ii}) is usually present in the early spectra of Type Ia SNe, either as PVF \citep{marion2014} or DF \citep{parrent2012}. \ion{Mg}{ii} may show more overlapping lines in the studied spectral range, but the red wing of the  $\lambda$4481 feature is relatively free of other blends, offering an excellent opportunity to measure its $v_{min}$. \ion{Mg}{ii} is a typical detached ion in the atmosphere of SN 2010kg at the earliest epochs, but its initially high minimum velocity decreases rapidly until it becomes permanently a PVF. This velocity profile is similar to the results from the analysis of the SN 2014J (categorized as a broad-line Ia, just like SN 2010kg) optical spectra by \cite{vallely2015} and \cite{marion2015}, where $v_{min}$ of \ion{Mg}{ii} declines through the epochs between -10 and +2 days. However, studying the near-infrared spectra of SN 2011fe \cite[categorized as a core normal subtype;][]{parrent2012}, \cite{hsiao2013} found that the \ion{Mg}{ii} velocity has a floor at $\sim$11,000 km s$^{-1}$ between -10 and +10 days, which is supposedly due to a bottom limit of \ion{Mg}{ii} abundance enhancement in SN 2011fe. Since \ion{Mg}{ii} is produced only in carbon burning, the existence of \ion{Mg}{ii} $v_{min}$ floor suggests the presence of an inner edge of carbon burning in SN 2011fe \citep{wheeler1998}. The lack of the \ion{Mg}{ii} velocity floor in SN 2010kg implies that either carbon burning may reach deeper layers in broad line SNe Ia than in core normal SNe Ia, or, alternatively, mixing might be more effective in the former type of ejecta.

\ion{Fe}{ii} and \ion{Fe}{iii} usually appear together in the supernova ejecta, because they produce relatively high optical depth around T $\sim$10,000 K \citep{hatano1999a}. Both \ion{Fe}{ii} and \ion{Fe}{iii} have several lines between 4100 and 5500 \r{A}, where the spectrum is strongly affected by \ion{Si}{ii} and \ion{S}{ii}. Although the minimum velocity of the iron ions are ambiguous, it is common to use \ion{Fe}{ii} as a detached and \ion{Fe}{iii} as a photospheric ion. In accord with these assumptions, we identify the lines of \ion{Fe}{iii} as PVFs and \ion{Fe}{ii} as DFs. The velocity of the latter one is $\sim$20,000 km s$^{-1}$ at the early epochs, which decreases linearly in time (see Fig. \ref{fig:fig4}). Unlike \ion{Mg}{ii}, the \ion{Fe}{ii} minimum velocity does not converge toward $v_{phot}$ during the observed temporal evolution of SN 2010kg.

\begin{figure*}
	% T\ion{O}{i}nclude a figure from a file named example.*
	% Allowable file formats are eps or ps if compiling using latex
	% or pdf, png, jpg if compiling using pdflatex
	\includegraphics[width=130mm]{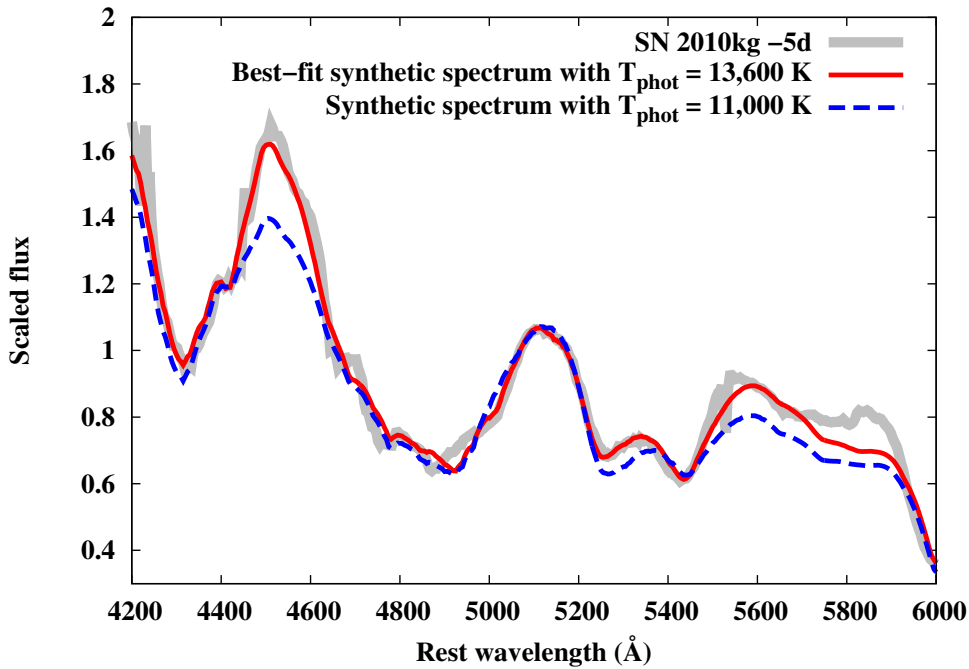}
    \caption{The measured spectrum 5 days before maximum light (gray), the best fit synthetic spectrum with $T_{phot}$ = 13,600 K (red), and a model spectrum having $T_{phot}$ = 11,000 K (blue). The lower temperature model spectrum produces an inferior fit around 4500 \r{A}. Above 6000 \r{A} the two models are indistinguishable. Note that the observed spectrum ends at 4100 \r{A} and the flux under 4300 \r{A} is uncertain.} 
    \label{fig:fig5}
\end{figure*}

The presence of \ion{S}{ii} affects a wide range between 4200 and 5200 \r{A}, but these lines are mostly weak. The two dominant absorption features are close to each other at $\lambda5463$ and $\lambda5641$ \r{A}, often called the \ion{S}{ii} "W" feature. The bluer absorption peak is overlapped by the emission part of the iron lines, but the redder peak is nearly free from other blending features. Because the line-forming region of \ion{S}{ii} begins right above the photosphere (at least before maximum light), thus \ion{S}{ii} $\lambda$5500 is ideal to test the $v_{phot}$ values. 

To fit the absorption feature around $\sim$5700 \r{A}, we adopted two more ions, \ion{Na}{i} and \ion{Ti}{ii}, because \ion{Si}{ii} $\lambda5973$ is not strong enough to fit the entire feature. The existence of both of these ions are possible in a Type Ia ejecta. The origin of titanium ions is usually explained by the thin helium-shell burning \citep{dessart2015, valenti2013} in the literature, which is one of the possible triggering mechanisms of the Type Ia explosions. \ion{Ti}{ii} is necessary only in the earliest spectra of SN 2010kg with minimum velocity values of 20 - 22,000 km s$^{-1}$. Our fitting uses \ion{Na}{i} as a detached ion that shows nearly constant velocity ($\sim$13,000 km s$^{-1}$) until maximum light. Note that the fit to the absorption feature around $\sim$5800 \r{A} is very degenerate, thus, we cannot fully confirm the presence of either \ion{Na}{i} or \ion{Ti}{ii}.

As noted in Section \ref{sec:ions}, both \ion{Ni}{ii} and \ion{Co}{ii} were also included in the initial models to reproduce the strong flux decrease below 4000 \r{A} as seen in every observed Type Ia SN spectrum \cite[e.g.][]{foley2012a, foley2012b}. This spectral region was not covered by the HET spectra, which prevents constraining the optical depths of the \ion{Ni}{ii} and \ion{Co}{ii} features. Therefore, we used these ions only to test whether their strong presence in the near-UV regime would affect the optical spectra. For this purpose we set the optical depths of \ion{Ni}{ii} and \ion{Co}{ii} by hand, and varied them until the shape of the spectrum below 4000 \r{A} looked like those of other SNe Ia. The minimum velocities for these ions were fixed at $v_{phot}$. We found that the contribution of these ions is minor, almost negligible in the optical spectrum above 4300 \r{A}. Between 4100 and 4300 \r{A} a pseudo-emission peak from the \ion{Co}{ii} features may appear, but since our flux calibration is rather uncertain in this region, we cannot use this part of the spectrum to get realistic constraints on the strength of these features. Thus, we decided to omit both \ion{Ni}{ii} and \ion{Co}{ii} from the final fits. Note that these spectrum models are valid only in the studied spectral range (4300 - 8500 \r{A}), and should  not be treated as a complete description of the chemical composition of the ejecta of SN 2010kg.

Although our SYN++ models show a decent fit to the spectra of SN 2010kg, an unfit feature appears at -1 day in the observed spectra at $\sim$5900 \r{A}, and it strengthens with time (see Fig.~\ref{fig:fig12}). Since there is no strong P Cygni feature in this area, whose pseudo-emission could explain the observed flux, it may originate from an emission source, which cannot be treated with SYN++. Such emission line is not expected in Type Ia supernova around maximum light. Based on their theoretical model, \cite{dessart2014a} showed that the forbidden transition of [\ion{Co}{iii}] $\lambda$5888 may be able to form such an emission line in delayed detonation Type Ia SNe. In their models the emission feature of [\ion{Co}{iii}] $\lambda$5888 appears at a few days after maximum light, a week later than our best-fit SYN++ models start to deviate from the observations at $\sim$5900 \r{A} in SN 2010kg.

\section{Discussion}
\label{sec:discussion}

\subsection{Temperature at the photosphere}
\label{sec:phottemperature}

The $T_{phot}$ values obtained from our SYN++ models ($\sim$13 - 15,000 K depending on epoch) are higher than the photospheric temperatures from some recent theoretical SN Ia models \cite[e.g.][]{kasen2006, jack2011, dessart2014c, kromer2009}, which are typically under 11000 K. However, the lower temperature ($T_{phot}$ < 11,000 K at maximum light) indicated by these models cannot fit the observed spectra of SN 2010kg, because the synthetic spectra do not reproduce the flux under 4700 \r{A} (see Fig. \ref{fig:fig5}). This region may also be influenced by the unknown in-host reddening (see Sec.~\ref{sec:dataset}). Since we corrected the observed spectra only for the Milky Way reddening, these corrected flux values may be only lower limits of the true fluxes. Thus, the disagreement between the observed and the low-temperature model spectra could be even more pronounced than it is suggested by Fig. \ref{fig:fig5}.

Furthermore, the photospheric temperatures predicted by various, recently published SN Ia models, spread over a wide range, and some of them predict higher values. For example, \cite{hachinger2013} suggested $T_{phot}$ $\sim$ 11 - 13,000 K (depending on epoch) from the analysis of the SN 2010kg-like broad-line SN 2010jn. \cite{dessart2014c} derived similar (T $\sim$ 12,000 K) temperature around maximum light in the velocity range of 10 - 16,000 km s$^{-1}$, where most of the emergent spectrum is formed. \cite{branch2005} applied $T_{phot}$ $\sim$ 13 - 14,000 K to fit the spectra of SN 1994D with SYNOW around and after maximum light.

On the other hand, there are known cases when such temperatures do not fit the observed SN spectra. For example, \cite{tanaka2008} modelled pre-maximum Ia spectra with a Monte-Carlo radiative transfer code, and found that $T_{phot}$ $\sim$ 17 - 18,000 K was necessary to get decent fits to spectra at 7 - 10 days pre-maximum. These are significantly higher temperatures than the "canonical" 11 - 13,000 K ones mentioned above, even though \cite{tanaka2008} found that such high temperatures are characteristics of low-velocity gradient (LVG) SNe, while high-velocity gradient (HVG) SNe, like SN 2010kg, tend to have lower $T_{phot}$. An extreme example for a SN Ia with high $T_{phot}$ is SN 2012dn \citep{chakradhari2014}, where $T_{phot}$ $\sim$ 20,000 K was found for the pre-maximum spectra.  

Another fact that further complicates the interpretation of our derived $T_{phot}$ values is that this parameter is very poorly constrained in a SYN++ model, especially for Type Ia SNe where there is no well-defined continuum in the optical. Thus, $T_{phot}$ given by SYN++ cannot be taken at face value as a representative of a physically realistic temperature at the photosphere. Instead, it should be considered only as a fitting parameter, which describes the shape of the spectrum.

Overall, it is concluded that even though our fits suggest higher $T_{phot}$ values ($\sim$13 - 15,000 K) than what is commonly adopted for SNe Ia ($\sim$11 - 13,000 K), these may not be directly related to the true temperatures around the photosphere. Based on more realistic SNe Ia models (see above), such high temperatures may not be unexpected, though.

\subsection{Temperature distribution in the atmosphere of SN 2010kg}
\label{sec:temperature}

Based on the calculations by \cite{hatano1999a}, we are able to create constraints on the excitation/ionization temperature distribution in the ejecta of SN 2010kg, considering on the optical depth values of the detected ions. As a first approximation, if we assume $T_{exc}$ $\sim$ $T_{local}$, then the presence of photospheric \ion{Si}{ii} and the absence of \ion{Si}{ii}I suggests an upper limit for the temperature near the photosphere, because the turning point between their optical depth functions, where \ion{Si}{ii}I starts to dominate over \ion{Si}{ii}, is at $\sim$11,000 K. The optical depth functions of the sulphur ions also point toward $T_{exc}$ $\sim$ $T_{local}$ $\sim$ 11,000 K. Similarly, the detection of doubly-ionized iron at photospheric velocity suggests a lower limit for $T_{local}$ as $\sim$12,000 K.

\begin{figure}
	% T\ion{O}{i}nclude a figure from a file named example.*
	% Allowable file formats are eps or ps if compiling using latex
	% or pdf, png, jpg if compiling using pdflatex
	\includegraphics[width=\columnwidth]{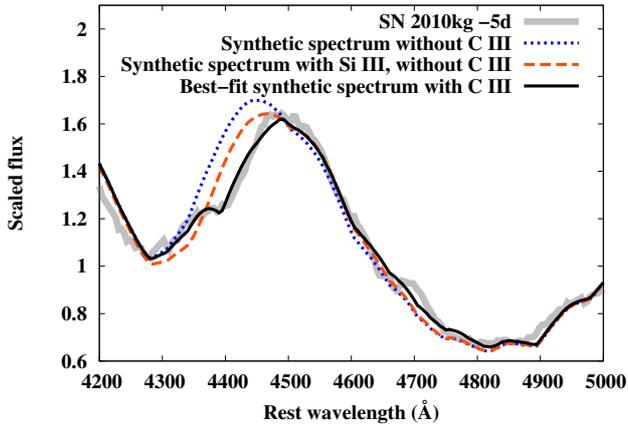}
    \caption{The absorption feature centered at $\sim$4400 \r{A} in the -5 days spectrum of SN 2010kg (grey line). The best-fit synthetic spectrum (including a \ion{C}{iii} detached line-forming region) is plotted with black solid line, the blue dashed line shows the synthesized spectrum without the \ion{C}{iii} contribution, while the orange dashed line shows the effect of replacing \ion{C}{iii} with photospheric \ion{Si}{ii}I. As seen, \ion{Si}{ii}I produces an inferior fit compared to \ion{C}{iii}.}
    \label{fig:fig6}
\end{figure}

\begin{figure}
	% T\ion{O}{i}nclude a figure from a file named example.*
	% Allowable file formats are eps or ps if compiling using latex
	% or pdf, png, jpg if compiling using pdflatex
	\includegraphics[width=\columnwidth]{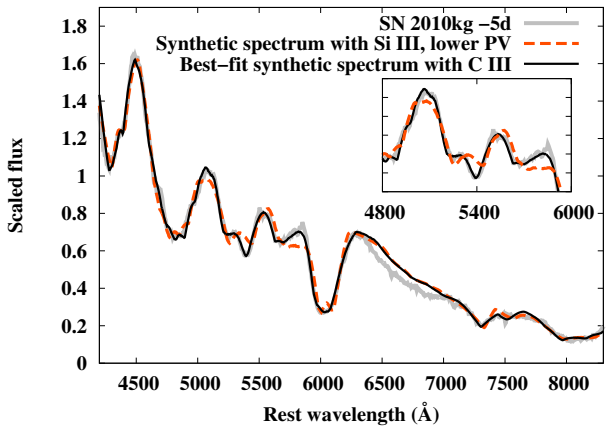}
    \caption{The measured spectrum 5 days before maximum light (gray), the best fit synthetic spectrum with $v_{phot}$ = 14,200 km s$^{-1}$ (black), and the effect of $v_{phot}$ = 11,000 km s$^{-1}$ needed for the well-fitting \ion{Si}{ii}I line (orange). Although the lower $v_{phot}$ value improves the fitting of the 4400 \r{A} feature with \ion{Si}{ii}I, the fit to the region around 5400 \r{A} gets worse, as also shown in the inset.}
    \label{fig:fig7}
\end{figure}

The estimated excitation temperature regime is in conflict with the $T_{phot}$ values in our best-fit SYN++ models. The reason for this discrepancy may be the same issue with the $T_{phot}$ parameter that we discussed in Section~\ref{sec:phottemperature}. Another probable reason is the lack of LTE in the atmosphere of SN 2010kg, as commonly found in other Type Ia SNe as well, which may make the excitation of different ions and species independent of the local temperature. Indeed, in a scattering-dominated SN ejecta, the color temperature of the emergent flux spectrum mimics the temperature of the deeper, thermalization layer, where true absorption starts to dominate over scattering \citep{jeffery1990}. Note that complete thermalization probably never occurs in Type Ia ejecta, thus, the emergent flux spectrum is not really Planckian \cite[see e.g.][]{branch2005}. As a result, $T_{exc}$ = $T_{phot}$ cannot be expected.

Proceeding toward the outer regions in the atmosphere of SN 2010kg, the local temperature may be estimable based on the model $T_{exc}$ values of the detached ions. The appearance of \ion{Fe}{ii} at $\sim$15 - 20,000 km s$^{-1}$ (depending on epoch) and the absence of detached \ion{Fe}{iii} suggests $T_{exc}$ < 10,000 K \citep{hatano1999a}. Based on our best-fit SYN++ models, $T_{local}$ $\sim$ 8 - 9,000 K is most likely for this region of the ejecta, because the optical depth of \ion{Fe}{ii} approaches 1 in this temperature regime. This estimate is also supported by the optical depth values of \ion{O}{i}, which stays as a detached ion through the pre-maximum epochs at velocities higher than \ion{Fe}{ii}.

The decrease of the local temperature from T $\sim$ 12,000 K to 8 - 9,000 K between v $\sim$ 10,000 km s$^{-1}$ and 20,000 km s$^{-1}$ is in accord with the predictions from recent delayed-detonation models \cite[e.g.][]{dessart2014c}.

\subsection{The feature at 4400 \r{A}}
\label{sec:oii}

The identification of the narrow absorption-like feature at $\sim$4400 \r{A} rest wavelength turned out to be an interesting issue that is worth investigating. Many authors fit this feature with \ion{Si}{iii} $\lambda4560$. Testing this assumption with the spectra of SN 2010kg, we find that \ion{Si}{iii} is unable to fit the red wing of this feature, if its velocity is set equal to $v_{phot}$ given by the fitting of \ion{S}{ii} and \ion{Fe}{iii} lines (see Fig. \ref{fig:fig6}). In order to get a decent fit to this feature by \ion{Si}{ii}I the only possibility is the decrease of $v_{phot}$ by $\sim$3000 km s$^{-1}$. However, this has an impact on the whole spectrum and disturbs the fitting of the \ion{Fe}{iii} $\lambda5156$, the \ion{S}{ii} W feature at $\sim$5300 \r{A} and the pseudo-emission of \ion{Si}{ii} $\lambda6355$ (see Fig. \ref{fig:fig7}). We suggest that the absorption feature centred at $\sim$4400 \r{A} is not likely formed by \ion{Si}{iii}, thus the existence of \ion{Si}{iii} in the ejecta is ambiguous. The only other feature which may support the presence of this ion (the small absorption notch at $\sim$5500 \r{A}) is not strong enough to characterize its line-forming region. It is still possible that the line at $\sim$4400 \r{A} is caused by the mixed contribution of \ion{Si}{iii} and other ions, but because of its high uncertainty, we omit \ion{Si}{iii} from the fitting. 

\begin{figure}
	% T\ion{O}{i}nclude a figure from a file named example.*
	% Allowable file formats are eps or ps if compiling using latex
	% or pdf, png, jpg if compiling using pdflatex
	\includegraphics[width=\columnwidth]{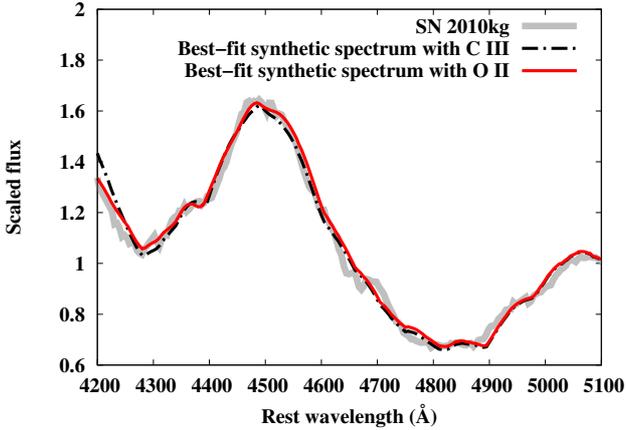}
    \caption{The absorption feature centered at $\sim$4400 \r{A} in the spectrum (grey line) of SN 2010kg, 5 days before B-max. The best fit synthetic spectrum (including a \ion{O}{ii} detached line-forming region) is plotted with the red solid line, while the black dashed line shows the synthesized spectrum with the \ion{C}{iii} contribution.}
    \label{fig:fig8}
\end{figure}

On the other hand, it is found that this feature can be fit much better by using either \ion{C}{iii}, or \ion{O}{ii}, as a detached ion (Fig. \ref{fig:fig8}). Previously, the presence of \ion{C}{iii} was identified in Type Ia spectrum only in the case of SN 1999aa \citep{garavini2004}, although, they found \ion{C}{iii} $\lambda$4647 as a PVF. \cite{chornock2006} also used \ion{C}{iii} to fit this feature in the spectrum of the Type Iax SN 2005hk, but they noted that this identification is rather ambiguous.

For 2010kg, the minimum velocity for either \ion{O}{ii} or \ion{C}{iii} must be $\sim$4 - 5000 km s$^{-1}$ higher than $v_{phot}$ to get a decent fit. If real, this would put these ions among the fastest ones (see Fig. \ref{fig:fig4}), but their line forming region stays definitely below that of the HVFs (see Fig. \ref{fig:fig3}). This is the region where remnants of the unburned matter of the original C/O white dwarf are expected. 

\begin{figure}
	% T\ion{O}{i}nclude a figure from a file named example.*
	% Allowable file formats are eps or ps if compiling using latex
	% or pdf, png, jpg if compiling using pdflatex
	\includegraphics[width=\columnwidth]{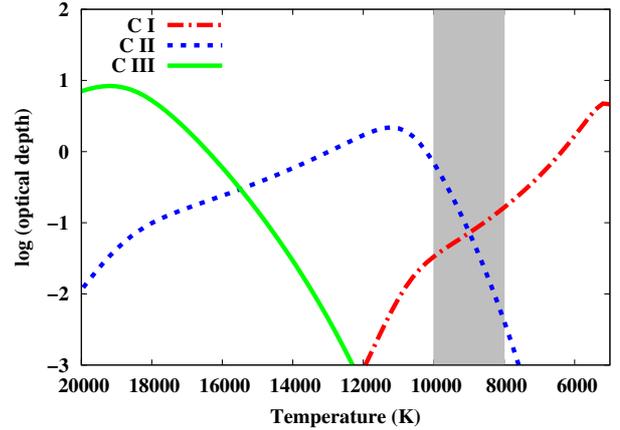}
    \caption{The solid lines show the optical depth values of carbon ions with electron density $N_{e} = 5 \times 10^{9}$ cm$^{-3}$ and number density of carbon $N_{i} = 5 \times 10^{9}$ cm$^{-3}$, after \protect\cite{hatano1999a}. The grey strip shows the local temperature range of the theoretical models in the outer part of Type Ia SN ejecta.}
    \label{fig:fig9}
\end{figure}

\begin{figure}
	% T\ion{O}{i}nclude a figure from a file named example.*
	% Allowable file formats are eps or ps if compiling using latex
	% or pdf, png, jpg if compiling using pdflatex
	\includegraphics[width=\columnwidth]{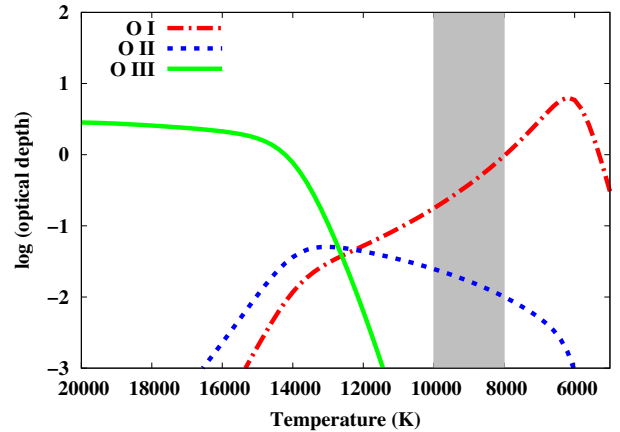}
    \caption{Optical depth values of oxygen ions with electron density $N_{e} = 5 \times 10^{9}$ cm$^{-3}$ and number density of oxygen $N_{i} = 5 \times 10^{9}$ cm$^{-3}$. The grey strip shows the local temperature range of the theoretical models in the outer part of Type Ia SN ejecta.}
    \label{fig:fig10}
\end{figure}

Since either \ion{C}{iii} $\lambda$4647 or \ion{O}{ii} $\lambda$4640 provides almost the same quality-of-fitting to the feature at 4400 \r{A}, we need to seek additional constraints to be able to identify this feature. We calculate the optical depths for different ions of carbon (Fig. \ref{fig:fig9}) and oxygen (Fig. \ref{fig:fig10}) as functions of the local temperature, similar to \cite{hatano1999a}. The grey shaded area represents the temperature range expected at the outermost ejecta region (Sec. \ref{sec:temperature}). It is seen that \ion{C}{iii} is unlikely to be really present in SN 2010kg, because this ion would require a local temperature well above 15,000 K. Instead, for temperatures lower than 10,000 K, it would be more feasible to find either \ion{C}{i} or \ion{C}{ii}, but there is no sign for any of these carbon ions in the optical spectra of SN 2010kg. All these pieces of evidence point toward the conclusion that the real presence of \ion{C}{iii} is not probable.

Although the "conventional" delayed-detonation models \cite[e.g.][]{kasen2006, jack2011, dessart2014c} usually predict a decreasing temperature profile toward the outer ejecta, at least for the pre-maximum epochs,  we also looked for alternatives, whether those might be able to support the appearance of \ion{C}{iii}. For example, some of the pulsating delayed-detonation models by \cite{dessart2014b} predict rising temperature profiles toward the outer region. Although the actual temperature values may depend on model assumptions, the range of the predicted outer temperatures is still not high enough to make the appearance of \ion{C}{iii} feasible.

One other possibility might be the significant decrease of the electron number density with respect to the value of $5 \times 10^9$ cm$^{-1}$ adopted by \cite{hatano1999a}, in the outer ejecta. Solving the Boltzmann- and Saha-equations one may find that this would move the peak of the \ion{C}{iii} ion optical depth toward lower temperatures. However, many orders of magnitude lower electron density would be necessary to get detectable amount of \ion{C}{iii} around T $\sim$ 10,000 K. Thus, we conclude that the presence of \ion{C}{iii} cannot be justified.

%\begin{figure}
	% T\ion{O}{i}nclude a figure from a file named example.*
	% Allowable file formats are eps or ps if compiling using latex
	% or pdf, png, jpg if compiling using pdflatex
%	\includegraphics[width=\columnwidth]{carbon_logtau}
%    \caption{Optical depth values of carbon ions with electron density $N_{e} = 5 \times 10^{9}$ cm$^{-3}$ and number density of carbon $N_{i} = 5 \times 10^{9}$ cm$^{-3}$, after \protect\cite{hatano1999a}. The grey strip
%shows the $T_{phot}$ range of the models at different epochs.}
%    \label{fig:fig8}
%\end{figure}

%\begin{figure}
	% T\ion{O}{i}nclude a figure from a file named example.*
	% Allowable file formats are eps or ps if compiling using latex
	% or pdf, png, jpg if compiling using pdflatex
%	\includegraphics[width=\columnwidth]{carbon_logtau2}
%    \caption{Optical depth values of carbon ions with electron density $N_{e} = 5 \times 10^{6}$ cm$^{-3}$ and number density of carbon $N_{i} = 5 \times 10^{6}$ cm$^{-3}$. The grey strip
%shows the $T_{phot}$ range of the models at different epochs.}
%    \label{fig:fig9}
%\end{figure}

If, instead, we assume that the 4400 \r{A} feature is due to \ion{O}{ii}, then we find that the minimum velocity for \ion{O}{ii} must be higher than $v_{phot}$ by 4 - 5,000 km s$^{-1}$ for every observed epoch. This is consistent with the \ion{O}{i} velocities that are also 6 - 7,000 km s$^{-1}$ above the photospheric velocity (see Fig. \ref{fig:fig4}). Since these two oxygen ions show the highest velocity from the ions forming DFs, they occur in the outer region of the SN 2010kg ejecta. Theoretical models \cite[e.g.][]{seitenzahl2013, moll2014}, which compute nuclear yield in Type Ia supernovae, also show high oxygen abundance in the outer region of the SN Ia atmosphere. In these models, the oxygen appears with nearly constant mass fraction in the ejecta down to $\sim$11 - 12,000 km s$^{-1}$.

In Fig. \ref{fig:fig10} we plot the optical depths of oxygen ions as functions of the temperature, again following the calculations by \cite{hatano1999a}. It is seen that the predicted optical depth values for \ion{O}{ii}, being between 0.01 and 0.1 in the grey-shaded temperature regime, are in good agreement with the values from our best-fit SYN++ models (see also the tables in the Appendix). Moreover, this is also true for the \ion{O}{i} optical depths found by SYN++. Thus, the simultaneous presence of \ion{O}{i} and \ion{O}{ii} in the outer part of the ejecta of SN 2010kg seems to be plausible.

Note that OI PVF is also detected in the spectra of SN 2010kg at 3 days before maximum light and afterward. This may raise the question: why does not the PVF component appear in the \ion{O}{ii} feature as well at these epochs? One possible reason might be the spatial separation of different oxygen-rich line forming regions in the ejecta. Oxygen can occur in Type Ia atmosphere either as unburnt material of the WD or as the result of fusion process \citep{maeda2010}. Because of their spatial separation, oxygen PVFs are more likely formed in the freshly synthesized oxygen layer, which is expected to be located deeper in the ejecta. Thus, the number density of oxygen could be significantly different at different spatial regimes in the ejecta. If we assume lower amount of oxygen around the photosphere (< $v_{phot}$ + 2000 km s$^{-1}$), only \ion{O}{i} may have high enough optical depth to produce a detectable optical line. This suggestion may be supported by the models of \cite{seitenzahl2013} and \cite{moll2014}, where the computed oxygen abundance drops one or two orders of magnitude (depending on the model) below $\sim$11,000 km s$^{-1}$, but remains significant even at $\sim$5,000 km s$^{-1}$.

\section{Conclusions}
\label{sec:conclusions}

We present eleven observed spectra of the Type Ia supernova SN 2010kg obtained between -10 and +5 days with respect to the B-band maximum, and their direct spectral analysis made via fitting synthetic spectra calculated with SYN++. The high cadence of the spectroscopic data give us the opportunity to reveal the velocity evolution of the identified ions. We distinguish the observed absorption features into three groups: high velocity features (HVFs), photospheric velocity features (PVFs) and detached features (DFs). 
The line-forming regions of HVFs show minimum velocities higher than 20,000 km s$^{-1}$ at the earliest epochs, and about three days before maximum light they settle at a constant velocity between 18,000 and 20,000 km s$^{-1}$. The typical HVFs are formed by \ion{Ca}{ii} and \ion{Si}{ii} in SN 2010kg. PVFs have (nearly) the same $v_{min}$ as the photosphere. The $v_{phot}$ is usually determined by fitting the \ion{S}{ii} and \ion{Fe}{iii} lines, but \ion{Si}{ii}, \ion{Ca}{ii}, \ion{O}{i}, and \ion{Mg}{ii} also form PVFs at certain observed epochs. Detached ions have $v_{min}$ > $v_{phot}$ but they are formed in the SN atmosphere, thus the minimum velocity of DFs stay below the $v_{HVF}$. Singly-ionized magnesium and iron ions start as detached ions at -10 days, but \ion{Mg}{ii} converges to the photosphere in time, while the \ion{Fe}{ii} line forming region stays always above the photosphere. \ion{O}{i} DF is detected with the highest velocity among the detached features. It might be formed in the unburned remnant of the progenitor white dwarf.

The deep, narrow feature at $\sim$4400 \r{A} is fit by \ion{C}{iii} $\lambda4647$ or \ion{O}{ii} $\lambda4650$ rather than \ion{Si}{ii}I $\lambda$4560. This attempt is slightly in tension with other spectral analysis results published recently, thus, we study the possibility of doubly-ionized carbon and singly-ionized oxygen ions in the SN ejecta. Calculating the temperature dependency of the optical depths for these ions, we find that high temperature (T $\sim$ > 16,000 K) is necessary to explain the presence of \ion{C}{iii} in the outer regions of the ejecta. Since recent theoretical models suggest that the temperature decreases toward the outer atmosphere in a Type Ia SNe, the presence of \ion{C}{iii} is highly unlikely.

Instead, the absorption feature at $\sim$4400 \r{A} could be more likely due to \ion{O}{ii}, which is able to produce a detectable optical feature at 7 - 9,000 K. This temperature regime is in good agreement with the temperature profile of SN 2010kg estimated from the optical depth functions of the detected ions. To our knowledge, this could be the first identification of ionized oxygen in a Type Ia SN atmosphere.

\section*{Acknowledgements}

This work has been supported by the Hungarian OTKA Grant NN107637 and NSF AST-1109801. J.M. Silverman is supported by an NSF Astronomy and Astrophysics Postdoctoral Fellowship under award AST-1302771.

The Hobby-Eberly Telescope (HET) is a joint project of the University of Texas at Austin, Stanford University, Ludwig-Maximilians-Universit\"{a}t M\"{u}nchen, and Georg-August-Universit\"{a}t G\"{o}ttingen. The HET is named in honor of its principal benefactors, William P. Hobby and Robert E. Eberly.
The Marcario Low Resolution Spectrograph is named for Mike Marcario of High Lonesome Optics who fabricated several optics for the instrument but died before its completion. The LRS is a joint project of the Hobby-Eberly Telescope partnership and the Instituto de Astronom\'{i}a de la Universidad Nacional Aut\'{o}noma de M\'{e}xico.

\appendix

\section{Observed spectra of the SN 2010kg, and the best fit synthetic spectra}
\label{sec:app}
The following tables contain the SYN++ model parameters, which were calculated assuming exponential density profile and a reference velocity of 13,000 km s$^{-1}$. $T_{phot}$ is the temperature of the photosphere, $v_{phot}$ is the velocity of the photosphere, $a0$ is the flux scaling parameter, while $\tau$ is the optical depth of the ion, aux is the e-folding parameter of its distribution, $v_{min}$ is the minimum velocity of the its line forming region and $T_{exc}$ the Boltzmann-excitation temperature. 

\begin{figure*}
	% T\ion{O}{i}nclude a figure from a file named example.*
	% Allowable file formats are eps or ps if compiling using latex
	% or pdf, png, jpg if compiling using pdflatex
	\includegraphics[width=160mm]{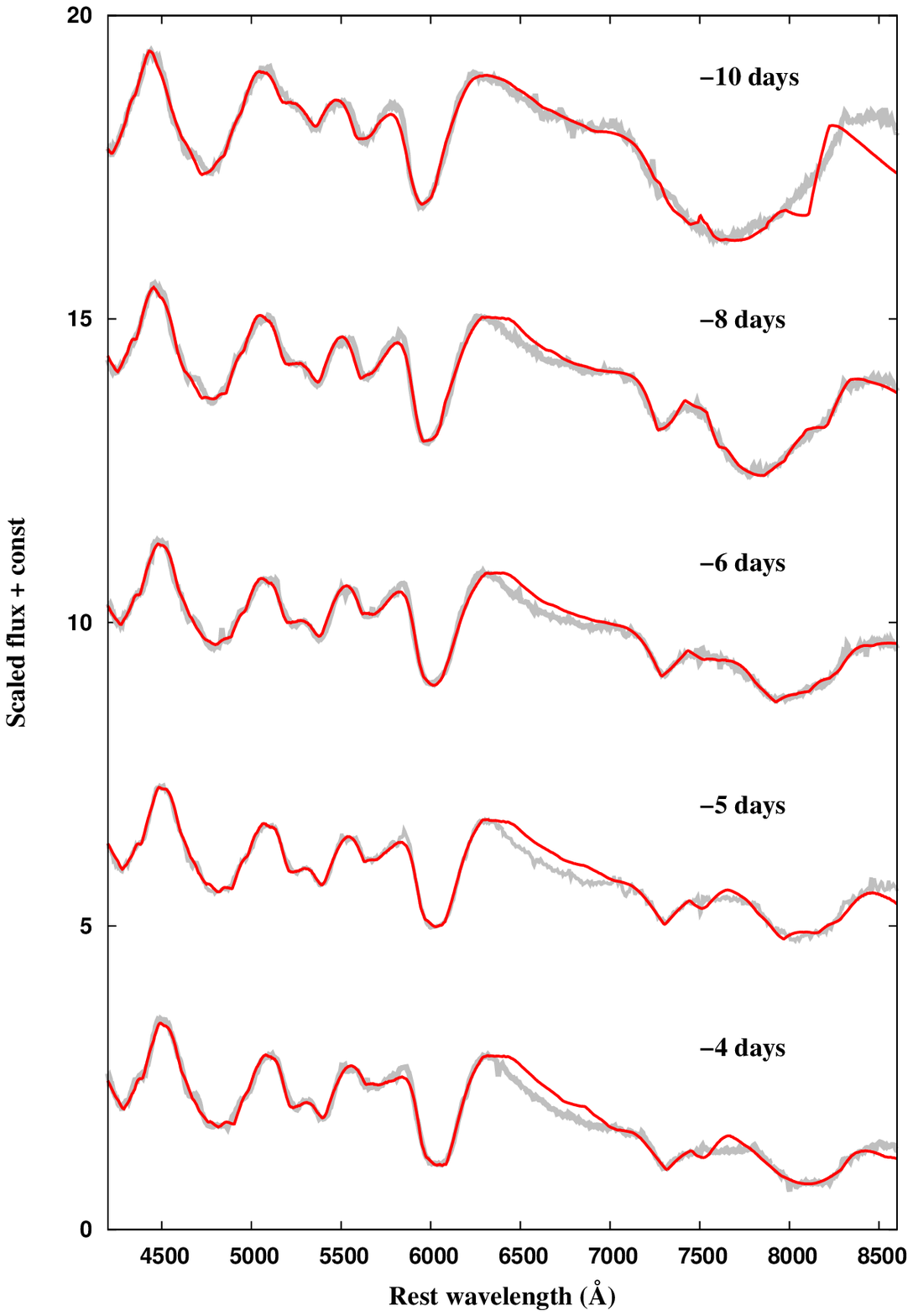}
    \caption{The observed spectra (grey lines) between epochs -10 and -4 days with respect to B-max, and the best fit synthetic spectra (red lines)}
    \label{fig:fig11}
\end{figure*}

\begin{figure*}
	% T\ion{O}{i}nclude a figure from a file named example.*
	% Allowable file formats are eps or ps if compiling using latex
	% or pdf, png, jpg if compiling using pdflatex
	\includegraphics[width=160mm]{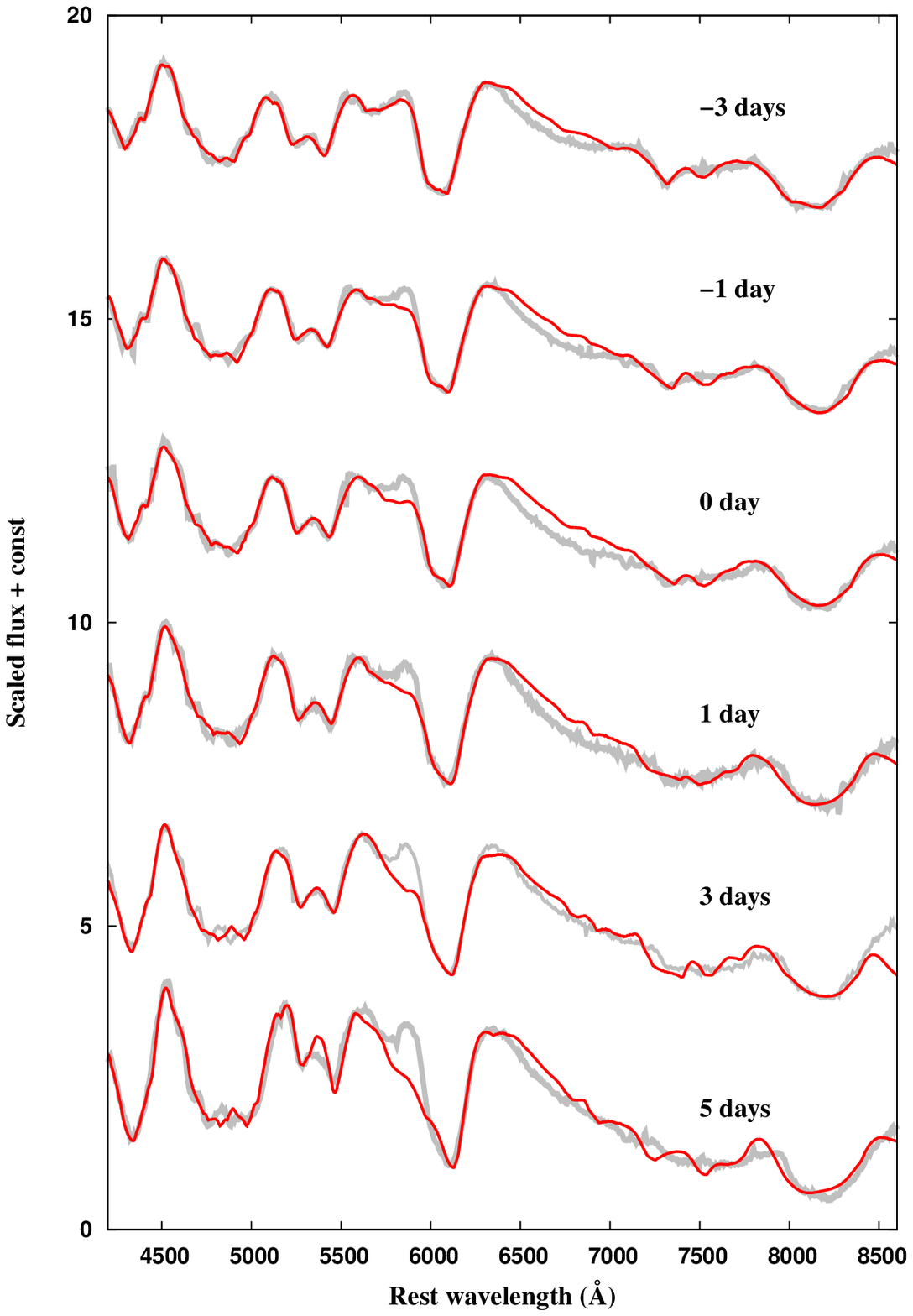}
    \caption{The observed spectra (grey lines) between epochs -3 and +5 days with respect to B-max, and the best fit synthetic spectra (red lines)}
    \label{fig:fig12}
\end{figure*}

\begin{table*}
	\centering
	\caption{The SYN++ parameters of the best-fit synthetic spectrum 10 days before maximum light. The $v_{min}$ values are in 10$^{3}$ km s$^{-1}$, while the $T_{exc}$ values are in 10$^{3}$ K. $a0$ = 0.77, $v_{phot}$ = 16.000 km s$_{-1}$ and $T_{phot}$ = 15.000 K. }
	\label{tab:example_table}
	\begin{tabular}{l|ccccccccccccc} % four columns, alignment for each
		\hline
		- & \ion{Ca}{ii}$_{PVF}$ & \ion{Ca}{ii}$_{HVF}$ & \ion{Si}{ii}$_{PVF}$ & \ion{Si}{ii}$_{HVF}$ & \ion{O}{i}$_{PVF}$ & \ion{O}{i}$_{DF}$ & \ion{O}{ii} & \ion{Mg}{ii} & \ion{S}{ii} & \ion{Fe}{ii} & \ion{Fe}{iii} & \ion{Na}{i} & \ion{Ti}{ii} \\
		\hline
		log(tau) & - & 4.20 & 1.00 & 3.35 & - & 0.70 & -0.60 & 0.58 & 0.07 & 0.30 & -0.10 & 0.07 & -0.40\\
		$v_{min}$ & - & 30.4 & 17.4 & 21.5 & - & 22.0 & 20.7 & 18.6 & 16.0 & 20.0 & 16.3 & 16.0 & 17.4\\
		aux & - & 5.50 & 3.80 & 1.67 & - & 3.50 & 1.62 & 5.50 & 2.35 & 9.40 & 1.40 & 5.00 & 3.00\\
		T$_{exc}$ & - & 14.6 & 10.5 & 10.5 & - & 5.0 & 16.0 & 6.0 & 11.5 & 5.0 & 5.0 & 10.0 & 20.4\\
		\hline
	\end{tabular}
\end{table*}

\begin{table*}
	\centering
	\caption{The SYN++ parameters of the best-fit synthetic spectrum 8 days before maximum light. The $v_{min}$ values are in 10$^{3}$ km s$^{-1}$, while the $T_{exc}$ values are in 10$^{3}$ K. $a0$ = 0.76, $v_{phot}$ = 15.200 km s$_{-1}$ and $T_{phot}$ = 14.600 K. }
	\label{tab:example_table}
	\begin{tabular}{l|ccccccccccccc} % four columns, alignment for each
		\hline
		- & \ion{Ca}{ii}$_{PVF}$ & \ion{Ca}{ii}$_{HVF}$ & \ion{Si}{ii}$_{PVF}$ & \ion{Si}{ii}$_{HVF}$ & \ion{O}{i}$_{PVF}$ & \ion{O}{i}$_{DF}$ & \ion{O}{ii} & \ion{Mg}{ii} & \ion{S}{ii} & \ion{Fe}{ii} & \ion{Fe}{iii} & \ion{Na}{i} & \ion{Ti}{ii} \\
		\hline
		log(tau) & 1.00	& 1.94 & 1.59 & 2.95 & - & 1.74 & -0.90 & 0.13 & 0.95 & 0.15 & -0.82 & 0.77 & -0.74\\
		$v_{min}$ & 17.5 & 26.2 & 15.8 & 21.1 & - & 21.1 & 20.1 & 16.0 & 15.2 & 19.3 & 15.6 & 15.2 & 16.7\\
		aux & 6.00 & 8.28 & 1.46 & 1.64 & - & 2.00 & 1.74 & 5.84 & 1.90 & 8.41 & 2.00 & 2.29 & 2.46\\
		T$_{exc}$ & 14.4 & 14.4 & 8.0 & 8.0 & - & 6.6 & 16.0 & 5.0 & 9.6 & 5.0 & 5.3 & 6.6 & 20.5\\
		\hline
	\end{tabular}
\end{table*}

\begin{table*}
	\centering
	\caption{The SYN++ parameters of the best-fit synthetic spectrum 6 days before maximum light. The $v_{min}$ values are in 10$^{3}$ km s$^{-1}$, while the $T_{exc}$ values are in 10$^{3}$ K. $a0$ = 0.71, $v_{phot}$ = 14.700 km s$_{-1}$ and $T_{phot}$ = 14.500 K. }
	\label{tab:example_table}
	\begin{tabular}{l|ccccccccccccc} % four columns, alignment for each
		\hline
		- & \ion{Ca}{ii}$_{PVF}$ & \ion{Ca}{ii}$_{HVF}$ & \ion{Si}{ii}$_{PVF}$ & \ion{Si}{ii}$_{HVF}$ & \ion{O}{i}$_{PVF}$ & \ion{O}{i}$_{DF}$ & \ion{O}{ii} & \ion{Mg}{ii} & \ion{S}{ii} & \ion{Fe}{ii} & \ion{Fe}{iii} & \ion{Na}{i} & \ion{Ti}{ii} \\
		\hline
		log(tau) & 1.26	& 1.85 & 1.37 & 2.52 & - & 0.77 & -1.07 & 0.06 & 0.80 & 0.05 & -0.46 & 0.50 & - \\\
		$v_{min}$ & 16.0 & 23.4 & 15.2 & 20.9 & - & 20.1 & 18.7 & 14.7 & 14.7 & 17.5 & 14.7 & 14.3 & - \\
		aux & 5.00 & 4.80 & 1.70 & 1.70 & - & 3.06 & 1.56 & 5.80 & 9.50 & 5.00 & 9.95 & 11.80 & - \\
		T$_{exc}$ & 15.5 & 15.5 & 7.2 & 7.2 & - & 6.3 & 15.9 & 13.0 & 9.5 & 5.0 & 10.0 & 11.8 & - \\
		\hline
	\end{tabular}
\end{table*}

\begin{table*}
	\centering
	\caption{The SYN++ parameters of the best-fit synthetic spectrum 5 days before maximum light. The $v_{min}$ values are in 10$^{3}$ km s$^{-1}$, while the $T_{exc}$ values are in 10$^{3}$ K. $a0$ = 0.68, $v_{phot}$ = 14.000 km s$_{-1}$ and $T_{phot}$ = 14.300 K. }
	\label{tab:example_table}
	\begin{tabular}{l|ccccccccccccc} % four columns, alignment for each
		\hline
		- & \ion{Ca}{ii}$_{PVF}$ & \ion{Ca}{ii}$_{HVF}$ & \ion{Si}{ii}$_{PVF}$ & \ion{Si}{ii}$_{HVF}$ & \ion{O}{i}$_{PVF}$ & \ion{O}{i}$_{DF}$ & \ion{O}{ii} & \ion{Mg}{ii} & \ion{S}{ii} & \ion{Fe}{ii} & \ion{Fe}{iii} & \ion{Na}{i} & \ion{Ti}{ii} \\
		\hline
		log(tau) & 1.50 & 2.25 & 1.46 & 2.02 & - & 0.28 & -1.26 & -0.12 & 0.70 & 0.27 & -0.36 & 0.30 & - \\
		$v_{min}$ & 14.0 & 21.8 & 14.6 & 20.4 & - & 19.3 & 18.0 & 14.2 & 14.0 & 17.0 & 14.0 & 14.0 & - \\
		aux & 3.90 & 3.32 & 1.47 & 1.95 & - & 4.80 & 1.54 & 6.20 & 1.65 & 4.30 & 2.00 & 2.30 & - \\
		T$_{exc}$ & 12.0 & 12.0 & 7.9 & 7.9 & - & 18.3 & 16.0 & 14.9 & 8.8 & 5.0 & 10.2 & 14.9 & - \\
		\hline
	\end{tabular}
\end{table*}

\begin{table*}
	\centering
	\caption{The SYN++ parameters of the best-fit synthetic spectrum 4 days before maximum light. The $v_{min}$ values are in 10$^{3}$ km s$^{-1}$, while the $T_{exc}$ values are in 10$^{3}$ K. $a0$ = 0.71, $v_{phot}$ = 13.600 km s$_{-1}$ and $T_{phot}$ = 14.100 K. }
	\label{tab:example_table}
	\begin{tabular}{l|ccccccccccccc} % four columns, alignment for each
		\hline
		- & \ion{Ca}{ii}$_{PVF}$ & \ion{Ca}{ii}$_{HVF}$ & \ion{Si}{ii}$_{PVF}$ & \ion{Si}{ii}$_{HVF}$ & \ion{O}{i}$_{PVF}$ & \ion{O}{i}$_{DF}$ & \ion{O}{ii} & \ion{Mg}{ii} & \ion{S}{ii} & \ion{Fe}{ii} & \ion{Fe}{iii} & \ion{Na}{i} & \ion{Ti}{ii} \\
		\hline
		log(tau) & 1.85 & 1.73 & 1.37 & 2.08 & - & 0.10 & -1.58 & -0.15 & 0.53 & 0.20 & -0.60 & -0.05 & - \\
		$v_{min}$ & 13.6 & 20.4 & 14.3 & 20.2 & - & 18.7 & 17.5 & 13.6 & 13.6 & 16.4 & 13.6 & 13.6 & - \\
		aux & 5.00 & 4.85 & 1.45 & 1.77 & - & 6.40 & 1.60 & 5.80 & 1.80 & 4.75 & 1.90 & 2.81 & - \\
		T$_{exc}$ & 12.9 & 12.9 & 6.4 & 6.4 & - & 25.0 & 16.2 & 24.4 & 8.0 & 5.0 & 7.7 & 5.0 & - \\
		\hline
	\end{tabular}
\end{table*}

\begin{table*}
	\centering
	\caption{The SYN++ parameters of the best-fit synthetic spectrum 3 days before maximum light. The $v_{min}$ values are in 10$^{3}$ km s$^{-1}$, while the $T_{exc}$ values are in 10$^{3}$ K. $a0$ = 0.68, $v_{phot}$ = 13.100 km s$_{-1}$ and $T_{phot}$ = 13.500 K. }
	\label{tab:example_table}
	\begin{tabular}{l|ccccccccccccc} % four columns, alignment for each
		\hline
		- & \ion{Ca}{ii}$_{PVF}$ & \ion{Ca}{ii}$_{HVF}$ & \ion{Si}{ii}$_{PVF}$ & \ion{Si}{ii}$_{HVF}$ & \ion{O}{i}$_{PVF}$ & \ion{O}{i}$_{DF}$ & \ion{O}{ii} & \ion{Mg}{ii} & \ion{S}{ii} & \ion{Fe}{ii} & \ion{Fe}{iii} & \ion{Na}{i} & \ion{Ti}{ii} \\
		\hline
		log(tau) & 1.45 & 1.95 & 1.11 & 2.28 & -0.15 & 0.08 & -1.67 & 0.20 & 0.49 & 0.16 & -0.83 & -0.12 & - \\
		$v_{min}$ & 13.1 & 20.0 & 13.9 & 19.8 & 13.1 & 18.6 & 16.8 & 13.1 & 13.1 & 16.1 & 13.1 & 13.6 & - \\
		aux & 4.20 & 3.40 & 1.49 & 1.50 & 1.29 & 5.13 & 1.55 & 2.34 & 1.74 & 4.60 & 2.64 & 2.81 & - \\
		T$_{exc}$ & 11.7 & 11.7 & 6.4 & 6.4 & 25.0 & 25.0 & 16.3 & 6.8 & 10.9 & 5.0 & 6.6 & 10.2 & - \\
		\hline
	\end{tabular}
\end{table*}

\begin{table*}
	\centering
	\caption{The SYN++ parameters of the best-fit synthetic spectrum 1 day before maximum light. The $v_{min}$ values are in 10$^{3}$ km s$^{-1}$, while the $T_{exc}$ values are in 10$^{3}$ K. $a0$ = 0.66, $v_{phot}$ = 12.200 km s$_{-1}$ and $T_{phot}$ = 13.700 K. }
	\label{tab:example_table}
	\begin{tabular}{l|ccccccccccccc} % four columns, alignment for each
		\hline
		- & \ion{Ca}{ii}$_{PVF}$ & \ion{Ca}{ii}$_{HVF}$ & \ion{Si}{ii}$_{PVF}$ & \ion{Si}{ii}$_{HVF}$ & \ion{O}{i}$_{PVF}$ & \ion{O}{i}$_{DF}$ & \ion{O}{ii} & \ion{Mg}{ii} & \ion{S}{ii} & \ion{Fe}{ii} & \ion{Fe}{iii} & \ion{Na}{i} & \ion{Ti}{ii} \\
		\hline
		log(tau) & 1.80 & - & 0.66 & 1.74 & 0.08 & -0.10 & -1.36 & -0.20 & 0.17 & 0.26 & -0.16 & -0.56 & - \\
		$v_{min}$ & 12.2 & - & 12.8 & 18.7 & 12.2 & 17.5 & 16.1 & 12.2 & 12.2 & 115.3 & 12.2 & 13.3 & - \\
		aux & 4.12 & - & 1.88 & 1.63 & 0.94 & 5.00 & 1.63 & 2.10 & 1.85 & 2.00 & 4.20 & 3.38 & - \\
		T$_{exc}$ & 9.8 & - & 7.0 & 7.0 & 25.0 & 25.0 & 14.8 & 10.6 & 9.2 & 5.0 & 9.7 & 12.0 & - \\
		\hline
	\end{tabular}
\end{table*}

\begin{table*}
	\centering
	\caption{The SYN++ parameters of the best-fit synthetic spectrum at maximum light. The $v_{min}$ values are in 10$^{3}$ km s$^{-1}$, while the $T_{exc}$ values are in 10$^{3}$ K. $a0$ = 0.68, $v_{phot}$ = 11.600 km s$_{-1}$ and $T_{phot}$ = 13.600 K. }
	\label{tab:example_table}
	\begin{tabular}{l|ccccccccccccc} % four columns, alignment for each
		\hline
		- & \ion{Ca}{ii}$_{PVF}$ & \ion{Ca}{ii}$_{HVF}$ & \ion{Si}{ii}$_{PVF}$ & \ion{Si}{ii}$_{HVF}$ & \ion{O}{i}$_{PVF}$ & \ion{O}{i}$_{DF}$ & \ion{O}{ii} & \ion{Mg}{ii} & \ion{S}{ii} & \ion{Fe}{ii} & \ion{Fe}{iii} & \ion{Na}{i} & \ion{Ti}{ii} \\
		\hline
		log(tau) & 1.80 & - & 0.54 & 1.50 & -0.10 & -0.06 & -2.13 & -0.48 & -0.02 & 0.15 & -0.48 & -0.90 & - \\
		$v_{min}$ & 11.6 & - & 12.2 & 18.0 & 11.6 & 16.8 & 15.70 & 11.6 & 11.6 & 15.2 & 11.6 & 13.0 & - \\
		aux & 4.00 & - & 1.92 & 1.65 & 1.70 & 6.60 & 1.61 & 2.50 & 1.80 & 4.10 & 0.50 & 4.50 & - \\
		T$_{exc}$ & 10.5 & - & 9.5 & 9.5 & 25.0 & 25.0 & 16.5 & 20.0 & 9.5 & 5.0 & 9.9 & 22.0 & - \\
		\hline
	\end{tabular}
\end{table*}

\begin{table*}
	\centering
	\caption{The SYN++ parameters of the best-fit synthetic spectrum 1 day after maximum light. The $v_{min}$ values are in 10$^{3}$ km s$^{-1}$, while the $T_{exc}$ values are in 10$^{3}$ K. $a0$ = 0.71, $v_{phot}$ = 11.000 km s$_{-1}$ and $T_{phot}$ = 13.500 K. }
	\label{tab:example_table}
	\begin{tabular}{l|ccccccccccccc} % four columns, alignment for each
		\hline
		- & \ion{Ca}{ii}$_{PVF}$ & \ion{Ca}{ii}$_{HVF}$ & \ion{Si}{ii}$_{PVF}$ & \ion{Si}{ii}$_{HVF}$ & \ion{O}{i}$_{PVF}$ & \ion{O}{i}$_{DF}$ & \ion{O}{ii} & \ion{Mg}{ii} & \ion{S}{ii} & \ion{Fe}{ii} & \ion{Fe}{iii} & \ion{Na}{i} & \ion{Ti}{ii} \\
		\hline
		log(tau) & 3.15 & - & 0.34 & 1.34 & 0.09 & -0.05 & -2.20 & -0.20 & -0.17 & 0.05 & -0.56 & -0.60 & - \\
		$v_{min}$ & 11.0 & - & 11.3 & 18.0 & 11.0 & 15.6 & 14.9 & 11.0 & 11.0 & 14.5 & 11.0 & 13.0 & - \\
		aux & 3.15 & - & 3.00 & 1.62 & 1.30 & 6.60 & 2.20 & 2.00 & 1.87 & 4.80 & 0.81 & 2.10 & - \\
		T$_{exc}$ & 10.7 & - & 7.6 & 7.6 & 25.0 & 25.0 & 15.6 & 9.4 & 8.7 & 5.0 & 10.0 & 25.0 & - \\
		\hline
	\end{tabular}
\end{table*}

\begin{table*}
	\centering
	\caption{The SYN++ parameters of the best-fit synthetic spectrum 3 days after maximum light. The $v_{min}$ values are in 10$^{3}$ km s$^{-1}$, while the $T_{exc}$ values are in 10$^{3}$ K. $a0$ = 0.70, $v_{phot}$ = 10.000 km s$_{-1}$ and $T_{phot}$ = 13.000 K. }
	\label{tab:example_table}
	\begin{tabular}{l|ccccccccccccc} % four columns, alignment for each
		\hline
		- & \ion{Ca}{ii}$_{PVF}$ & \ion{Ca}{ii}$_{HVF}$ & \ion{Si}{ii}$_{PVF}$ & \ion{Si}{ii}$_{HVF}$ & \ion{O}{i}$_{PVF}$ & \ion{O}{i}$_{DF}$ & \ion{O}{ii} & \ion{Mg}{ii} & \ion{S}{ii} & \ion{Fe}{ii} & \ion{Fe}{iii} & \ion{Na}{i} & \ion{Ti}{ii} \\
		\hline
		log(tau) & 2.03 & - & 0.09 & 1.25 & -0.40 & -0.02 & -2.00 & -0.30 & -0.35 & 0.00 & -1.60 & - & - \\
		$v_{min}$ & 10.0 & - & 11.1 & 18.0 & 10.0 & 15.0 & 14.3 & 10.0 & 10.0 & 13.0 & 10.0 & - & - \\
		aux & 2.00 & - & 3.91 & 1.42 & 1.00 & 6.00 & 2.31 & 1.00 & 2.13 & 6.38 & 0.50 & - & - \\
		T$_{exc}$ & 12.7 & - & 5.8 & 5.8 & 25.0 & 25.0 & 14.7 & 5.0 & 9.5 & 5.4 & 9.5 & - & - \\
		\hline
	\end{tabular}
\end{table*}

\begin{table*}
	\centering
	\caption{The SYN++ parameters of the best-fit synthetic spectrum 5 days after maximum light. The $v_{min}$ values are in 10$^{3}$ km s$^{-1}$, while the $T_{exc}$ values are in 10$^{3}$ K. $a0$ = 0.76,  $v_{phot}$ = 9.400 km s$_{-1}$ and $T_{phot}$ = 13.000 K. }
	\label{tab:example_table}
	\begin{tabular}{l|ccccccccccccc} % four columns, alignment for each
		\hline
		- & \ion{Ca}{ii}$_{PVF}$ & \ion{Ca}{ii}$_{HVF}$ & \ion{Si}{ii}$_{PVF}$ & \ion{Si}{ii}$_{HVF}$ & \ion{O}{i}$_{PVF}$ & \ion{O}{i}$_{DF}$ & \ion{O}{ii} & \ion{Mg}{ii} & \ion{S}{ii} & \ion{Fe}{ii} & \ion{Fe}{iii} & \ion{Na}{i} & \ion{Ti}{ii} \\
		\hline
		log(tau) & 2.00 & - & 0.12 & 1.15 & -0.06 & -0.29 & -1.98 & -0.30 & -0.75 & 0.00 & -1.50 & - & - \\
		$v_{min}$ & 9.4 & - & 10.8 & 17.5 & 9.4 & 14.4 & 14.0 & 9.4 & 9.4 & 11.9 & 9.4 & - & - \\
		aux & 3.90 & - & 3.11 & 1.50 & 2.50 & 9.50 & 3.20 & 1.50 & 0.85 & 6.16 & 0.66 & - & - \\
		T$_{exc}$ & 10.6 & - & 6.0 & 6.0 & 25.0 & 25.0 & 14.2 & 12.0 & 10.0 & 8.1 & 9.8 & - & - \\
		\hline
	\end{tabular}
\end{table*}
%%%%%%%%%%%%%%%%%%%%%%%%%%%%%%%%%%%%%%%%%%%%%%%%%%

% Don't change these lines
\bsp	% typesetting comment
\label{lastpage}
\end{document}